\documentclass{sig-alternate}

\usepackage{color}
\usepackage{url}
\usepackage{verbatim} % allows multiline comments
\usepackage{subfigure} % for the subfloat in the figs
\usepackage{multirow}
\usepackage{algorithm}
\usepackage[noend]{algorithmic}
\usepackage{xspace}
\usepackage{graphicx}
\usepackage{epsfig}
\usepackage{amsmath}
\usepackage{amssymb}
\usepackage{ctable}
\usepackage{times}
\usepackage{mathptmx}

\newcommand{\hide}[1]{}
\newcommand{\xhdr}[1]{\vspace{1.7mm}\noindent{{\bf #1.}}}
\newcommand{\casc}{{\mathbf{t}}}
\newcommand{\alphs}{{\mathbf{A}}}

\newtheorem{theorem}{Theorem}

\newcommand{\netinf}{{\textsc{Net\-Inf}}\xspace}
\newcommand{\multitree}{{\textsc{Multi\-Tree}}\xspace}
\newcommand{\netrate}{{\textsc{Net\-Rate}}\xspace}
\newcommand{\connie}{{\textsc{Co\-nNIe}}\xspace}

\newcommand{\infopath}{{\textsc{InfoPath}}\xspace}
\newcommand{\expo}{{\textsc{Exp}}\xspace}
\newcommand{\pow}{{\textsc{Pow}}\xspace}
\newcommand{\ray}{{\textsc{Ray}}\xspace}
\newcommand{\eg}{\emph{e.g.}}
\newcommand{\ie}{\emph{i.e.}}
\newcommand{\unobs}{{\infty}}

\graphicspath{{./FIG/}}

\begin{document}

\conferenceinfo{WSDM'13,} {February 4--8, 2013, Rome, Italy.} 
\CopyrightYear{2013} 
\crdata{978-1-4503-1869-3/13/02} 
\clubpenalty=10000 
\widowpenalty = 50000

\title{Structure and Dynamics of Information\\ Pathways in Online Media}

\numberofauthors{3}
\author{
\alignauthor Manuel Gomez-Rodriguez\\
        \affaddr{MPI for Intelligent Systems\\ Stanford University} \\
        \email{manuelgr@stanford.edu} \\
\alignauthor Jure Leskovec\\
        \affaddr{Stanford University} \\
        \email{jure@cs.stanford.edu} \\
\alignauthor Bernhard Sch\"{o}lkopf\\
        \affaddr{MPI for Intelligent Systems} \\
        \email{bs@tuebingen.mpg.de} \\
}

\maketitle

\begin{abstract}
% !TEX root = dynamic-network-inference.tex
Diffusion of information, spread of rumors and infectious diseases are all instances of stochastic processes that occur over the edges of an underlying network. Many times networks over
which contagions spread are unobserved, and such networks are often dynamic and change over time.
In this paper, we investigate the problem of inferring dynamic networks based on information diffusion data. We assume there is an unobserved dynamic network that changes over time,
while we observe the results of a dynamic process spreading over the edges of the network. The task then is to infer the edges and the dynamics of the underlying network.

We develop an on-line algorithm that relies on stochastic convex optimization to efficiently solve the dynamic network inference problem. We apply our algorithm to information diffusion
among 3.3 million mainstream media and blog sites and experiment with more than 179 million different pieces of information spreading over the network in a one year period. We study
the evolution of information pathways in the online media space and find interesting insights. Information pathways for general recurrent topics are more stable across time than for on-going 
news events. Clusters of news media sites and blogs often emerge and vanish in matter of days for on-going news events. Major social movements and events involving civil population, 
such as the Libyan'{}s civil war or Syria'{}s uprise, lead to an increased amount of information pathways among blogs as well as in the overall increase in the network centrality of blogs and 
social media sites.
\end{abstract}

\vspace{1mm}
\noindent {\bf Categories and Subject Descriptors:} H.2.8 {\bf
[Database Management]}: Database applications---{\it Data mining}

\noindent {\bf General Terms:} Algorithms; Experimentation.


\section{Introduction}
\label{sec:intro}
% !TEX root = dynamic-network-inference.tex
Networks represent a fundamental medium for spreading and diffusion of various types of behavior, information, rumors and diseases~\cite{rogers95diffusion}. A {\em contagion} appears 
at some node of a network and then spreads like an epidemic from node to node over the edges of the underlying network. For example, in case of information diffusion, the contagion 
represents a piece of information~\cite{leskovec2009kdd,nowell08letter} and infection events correspond to times when nodes mention or copy the information from one of their neighbors 
in the network. Similarly, we can think about the spread of a new type of behavior or an action, e.g., purchasing a new cellphone~\cite{jure06viral}, or the propagation of a contagious 
disease over the edges of the underlying social network~\cite{bailey75mathematical}.

%Recently there has been an increasing effort to computationally study and model information propagation in social networks, social media and the Web~\cite{aral2012, chierichetti2011, kempe03maximizing, leskovec2009kdd, ugander2012structural}.

% DESCRIPTION OF THE PROBLEM
In the context of network diffusion, we often observe the temporal traces of diffusion while the pathways over which contagion spreads remain hidden. In other words, we observe the times when each node gets infected by the contagion, but the edges of the network that gave rise to the diffusion remain unobservable.
For example, we can often measure and observe the time when people decide to adopt a new behavior while we do not explicitly observe which neighbor in the social network influenced them 
to do so. In case of information diffusion, we often observe people (or media sites) talking about a new piece of information without explicitly observing the path it took in the information diffusion 
network to reach the particular node of interest. And, epidemiologists often observe when a person gets sick but usually cannot tell who infected her. In all these examples, one can observe the 
infection events themselves while not knowing over which edges of the network the contagions spread. Therefore, one of the fundamental research problems in the context of network diffusion 
is inferring the structure of networks over which various types of contagions spread~\cite{manuel10netinf}. Moreover, many times networks over which contagions diffuse are not static but 
change over time. Depending on the type of contagion, the time of the day, or death of the existing and birth of new nodes, the underlying network may dynamically change 
and shift over time.

In recent years, several network inference algorithms have been developed~\cite{manuel11icml, manuel10netinf, multitree12icml, meyers10netinf, netrapalli12, snowsill2011refining}. Some approaches infer only the network structure~\cite{manuel10netinf, snowsill2011refining}, while others infer not only the network structure but also the \emph{strength} or the average latency of every
edge in the network~\cite{manuel11icml, meyers10netinf}. However, to the best of our knowledge, previous work has always assumed networks to be static and contagion pathways to be constant over time. However, in most cases, networks are dynamic, and contagion pathways change over time, depending upon the contagions that propagate through them~\cite{seth2012kdd,romero11twitter}.
For example, a blog can increase its popularity abruptly after one of its posts turns \emph{viral}, this may create new edges in the information transmission network and so the content the blog produces in the future will likely spread to larger parts of the network. Similarly, at any given time a particular unexpected event may occur and a topic or piece of news may become very popular for a limited period of time. This again will lead to different emerging and vanishing information pathways, and thus to a time-varying underlying network. In order to better understand these temporal changes, one needs to reconstruct the time-varying structure and underlying temporal dynamics of these networks and then study the information pathways of real-world events, topics or content.

\begin{table}[t]
    \begin{center}
    \begin{small}
    \begin{tabular*}{0.48\textwidth}{@{\extracolsep{\fill}} l  l l}
        \toprule
    %    \multirow{2}{*}{\textbf{Model}}
		%& \multicolumn{2}{l}{\textbf{Transmission likelihood}} \vspace{1mm} \\
		%& \multicolumn{2}{l}{$f(t_i|t_j;\alpha_{j,i})$} \\
		\textbf{Model} & \multicolumn{2}{l}{\textbf{Edge $\bf (j,i)$ transmission likelihood $\bf f(t_i|t_j;\alpha_{j,i})$}} \\
	\midrule
	\underline{\expo}{onential} & $\left\{
	\begin{array}{l l}
    	\alpha_{j,i}\cdot e^{-\alpha_{j,i} (t_i-t_j)} \\
 		0
	\end{array}\right.$ &
	\hspace{-8mm}
	$\begin{array}{l}
		\text{if $t_j < t_i$}\\
		\text{otherwise}
	\end{array}$ \\
	\midrule
	\underline{\pow}{er law} & $\left\{
	\begin{array}{l}
    	\frac{\alpha_{j,i}}{\delta} \left(\frac{t_i-t_j}{\delta}\right)^{-1-\alpha_{j,i}}\\
		0
	\end{array}\right.$ &
	\hspace{-8mm}
	$\begin{array}{l}
		\text{if $t_j+\delta < t_i$}\\
		\text{otherwise}\\
	\end{array}$
	\\
	\midrule
	\underline{\ray}{leigh} & $\left\{
	\begin{array}{l}
    	\alpha_{j,i} (t_i-t_j) e^{-\frac{1}{2} \alpha_{j,i} (t_i-t_j)^2}\\
		0
	\end{array}\right.$ &
	\hspace{-8mm}
	$\begin{array}{l}
		\text{if $t_j < t_i$}\\
		\text{otherwise}
	\end{array}$
	\\
	\bottomrule
    \end{tabular*}
    \end{small}
    \end{center}
    \vspace{-4mm}
    \caption{Various models of edge transmission likelihood.}
    \label{tab:likelihoods}
    \vspace{-5mm}
\end{table}

% OUR SOLUTION
\xhdr{Our approach to time-varying network inference}
In this paper we investigate the problem of inferring dynamic networks based on information diffusion data. We assume there is an unobserved dynamic network that changes over time, while we observe 
the node infection times of many different contagions spreading over the edges of the network. The task then is to infer the edges and the dynamics of the underlying network. We develop an efficient 
on-line dynamic network inference algorithm, \infopath, that allows us to infer daily networks of information diffusion between online media sites over a one year period using more than 179 million 
different contagions diffusing over the underlying media network.

We model diffusion processes as discrete networks of fully continuous temporal processes occurring at different rates building on our previous work~\cite{manuel11icml, influmax12icml}. Our model 
allows information to propagate at different rates across different edges by adopting a data-driven approach, where only the recorded temporal diffusion events are used. The model considers the information 
which propagates through the network due only to diffusion, while ignoring any ex\-ter\-nal sour\-ces~\cite{seth2012kdd}.
However, our original diffusion model considered only static networks~\cite{manuel11icml}. Here, we generalize the model and develop a new inference method to support dynamic networks. Our time-varying network inference algorithm, \infopath, uses stochastic gradient~\cite{robbins1951stochastic} to provide estimates of the time-varying structure and temporal dynamics of the inferred network. The framework enables us to study the temporal evolution of information pathways in the online media space.

% SUMMARIZE RESULTS IN SYNTHETIC AND REAL DATA (here or after related work?)
We apply the \infopath algorithm to synthetic as well as real Web information propagation data. We study 179 million different information cascades spreading among 3.3 million blog and news media sites over a
one year period, from March 2011 till February 2012.\footnote{All data, code and additional results are available at the supporting
website~\cite{website13}: \url{http://snap.stanford.edu/infopath}.}
Results on synthetic data show \infopath is able to track changes in the topology of dynamic networks and provides accurate on-line estimates of the time-varying transmission rates of the edges of the network. \infopath is also robust across network topologies, and temporal trends of edge transmission dynamics.% of transmission rate.

Experiments on large-scale real news and social media data lead to interesting insights and findings. For example, we find that the information pathways over which general recurrent topics propagate remain more stable over time, while unexpected events lead to dramatically changing information pathways.
Clusters of mainstream news and blogs often emerge and vanish in a matter of days, and our on-line algorithm is able to uncover such structures.
News events that involve large-scale social movements, as the Libyan civil war, Egypt'{}s revolution or Syria'{}s uprise, result in a greater increase in information transfer among blogs than among mainstream media.
Perhaps surprisingly, the amount of mainstream media and blogs among the most influential nodes for most topics or news events are comparable. However, we find that growing numbers of influential blogs on some topics or news events are often temporally correlated with large-scale social movements (\eg, the Occupy Wall Street movement in Sept-Nov 2011).

%% RELATED WORK
\xhdr{Further related work}
Previous methods for inferring diffusion networks~\cite{manuel11icml, manuel10netinf, multitree12icml, meyers10netinf} also use a generative probabilistic model
for modeling cascading processes over networks. \netinf~\cite{manuel10netinf} and \multitree~\cite{multitree12icml} infer the network connectivity using submodular
optimization. \netrate~\cite{manuel11icml} and \connie~\cite{meyers10netinf} infer not only the network connectivity but also transmission rates of infection or prior probabilities of infection using 
convex optimization. Moreover, there have been also attempts to model information diffusion without assuming the existence of an underlying network~\cite{yang2010patterns, yang2010}.
\begin{table*}[t]
    \small
    \begin{center}
    \begin{tabular*}{\textwidth}{@{\extracolsep{\fill}} l c c c c}
	\toprule
        \multirow{2}{*}{\textbf{Model}}
		& \textbf{Log survival function} & \textbf{Hazard function} & \textbf{Cascade gradient for uninfected} & \textbf{Cascade gradient for infected} \vspace{1mm} \\
		& $\log S(t_i | t_j ; \alpha_{j,i})$ & $H(t_i | t_j ; \alpha_{j,i})$ & $\nabla_{\alpha_{j,i}} L_{c}(\alphs)$ & $\nabla_{\alpha_{j,i}} L_{c}(\alphs)$ \\
	\midrule
	\expo & $-\alpha_{j,i} (t_i-t_j)$ & $\alpha_{j,i}$ & $T - t^c_j$ & $(t^c_i - t^c_j) - \frac{1}{\sum_{k : t^c_k < t^c_i} \alpha_{k,i}}$ \\
	\midrule
	\pow & $-\alpha_{j,i} \log\left(\frac{t_i-t_j}{\delta}\right)$ & $\alpha_{j,i}\cdot\frac{1}{t_i-t_j}$ & $\log \left(\frac{T - t^c_j}{\delta}\right)$ & $\log \left(\frac{t^c_i - t^c_j}{\delta}\right) - \frac{(t^c_i - t^c_j)^{-1}}{\sum_{k : t^c_k < t^c_i} \alpha_{k,i} (t^c_i - t^c_k)^{-1}}$ \\
	\midrule
	\ray & $-\alpha_{j,i} \frac{(t_i-t_j)^2}{2}$ & $\alpha_{j,i} \cdot (t_i-t_j)$ & $\frac{(T - t^c_j)^{2}}{2}$ & $\frac{(t^c_i - t^c_j)^{2}}{2} - \frac{t^c_i - t^c_j}{\sum_{k : t^c_k < t^c_i} \alpha_{k,i} (t^c_i - t^c_k)}$\\
	\bottomrule
    \end{tabular*}
    \end{center}
   \vspace{-5mm}
   \caption{Contagion transmission models for the three edge transmission likelihoods: Exponential, Power-law and Rayleigh.}
    \label{tab:log-likelihood-gradients}
    \vspace{-4mm}
\end{table*}

% WHY WHAT WE DO CAN HELP
However, to the best of our knowledge, all previous approaches to network inference assume the network and the underlying dynamics of the edges to be constant, \ie, the network structure and 
the transmission rates of each edge do not change over time. Therefore, they consider the pathways over which information propagates to be time-invariant. The main contribution of this paper is to 
combine stochastic gradient and the diffusion model introduced in~\cite{manuel11icml} to develop an efficient on-line network inference algorithm that provides time-varying estimates of the edges of 
a network and the transmission rates of each edge. This allows us to detect how information pathways emerge and vanish over time, and identify when nodes produce highly viral content.

% OUTLINE OF THE PAPER
The remainder of the paper is organized as follows: in Sec.~\ref{sec:formulation}, we revisit the model of diffusion and state the dynamic network inference problem.
Section~\ref{sec:proposed} describes the proposed time-varying network inference method, called \infopath. Section~\ref{sec:evaluation} evaluates \infopath quantitatively and qualitatively 
using synthetic and real diffusion data. We conclude with a discussion of results in Section~\ref{sec:conclusions}. 

\section{Problem formulation}
\label{sec:formulation}
% !TEX root = dynamic-network-inference.tex
In this section, we build on our fully continuous time model of diffusion~\cite{manuel11icml, influmax12icml}.
We start by briefly describing the generative model for the observed data. % and then in the next section describe the network inference method.
We then revisit how to compute the likelihood of a cascade using the model and state the continuous time network inference problem for both static and dynamic networks.
Across the section, we explicitly point out which assumptions of the original model need to be extended in order to support dynamic networks.

\xhdr{Observed data}
For now let's consider a single static directed network. Over the edges of the network multiple contagions
% (independently of each other) manuel: here we may like not to talk about independence, we state it in the next paragraph
propagate. As the contagion spreads from infected to non-infected nodes over the edges of the network the contagion creates a {\em cascade}. For each contagion $c$, we observe a cascade $\casc^c$, which is simply a record of observed node infection times during a time window of length $T^c$. In an information propagation setting, each cascade corresponds to a different piece of information and the infection time of a node is simply the time when the node first mentioned the piece of information $c$.

Cascade is a $N$-dimensional vec\-tor $\casc^c:=(t^c_1,\ldots,t^c_N)$ re\-cor\-ding the times when each of $N$ nodes got infected by the contagion $c$: $t^c_k\in [t_{0},t_{0}+T^c]\cup\{\unobs\}$, where
$t_{0}$ is the infection time of the first node. Generally contagions do not infect all the nodes of the network, symbol $\unobs$ is used for nodes that were not infected by the contagion $c$ during the 
observation window $[t_{0},t_{0}+T^c]$. Contagions often propagate simultaneously~\cite{myers12clash, prakash2012winner} over the same network but we assume each contagion to propagate 
independently of each other.

Given a set of node infection times of many different contagions, our goal is to infer the underlying dynamic network over which contagions propagated. We apply the Maximum Likelihood principle in order to infer the network that most likely generated the observed data. We proceed by assuming a static network and describe the generative model of information diffusion. We then generalize the model to dynamic networks.

\xhdr{Pairwise transmission likelihood} The first step in modeling diffusion dynamics is to consider pairwise node interaction. For eve\-ry pair of nodes $(j, i)$, we define a pairwise transmission rate $\alpha_{j,i}$ which models how frequently information spreads from node $j$ to node $i$; the \emph{strength} of an edge $(j, i)$. We pay attention to the rather general case of heterogeneous pairwise transmission rates, \ie, infections can occur at di\-ffe\-rent transmission rates over di\-ffe\-rent edges of the network.
As $\alpha_{j,i} \rightarrow 0$ the expected transmission time from node $j$ to node $i$ becomes arbitrarily long.
In contrast with the original model~\cite{manuel11icml}, we will later allow transmission rates $\alpha_{j,i}$ to change over time. In particular, we will allow the transmission rates $\alpha_{j,i}$ to change across cascades but not within a cascade. Allowing edge transmission rates to dynamically increase and decay over time will enable us to infer time-varying diffusion networks.

Next, we define $f(t_i | t_j ; \alpha_{j,i})$ to be the conditional likelihood of transmission between node $j$ and node $i$ and assume it depends on the
infection times $(t_j, t_i)$ and the edge transmission rate $\alpha_{j,i}$. We allow information to only propagate forward in time, i.e.,
node $j$ that has been infected at time $t_j$ may infect node $i$ at time $t_i$ only if $t_j < t_i$, otherwise $f(t_i | t_j, \alpha_{j,i}) = 0$.
The shape of the conditional likelihood of transmission may depend on the particular setting (information, influence, diseases, etc.) in which
propagation takes place. In some scenarios, it may be possible to estimate a non-parametric likelihood while in others, expert knowledge may be used to
decide upon a parametric model. For simplicity, we consider three well-known parametric models of edge transmission rates: exponential,  power-law, and Rayleigh, defined in
Table~\ref{tab:likelihoods}.
Exponential and power-law likelihoods have been used in modeling information propagation in social and information networks~\cite{manuel11icml, manuel10netinf,
influmax12icml, multitree12icml, meyers10netinf}, while Rayleigh has been used in previous work in disease spread in epidemiology~\cite{wallinga04epidemic}.
In all three models, as $\alpha_{j,i} \rightarrow 0$, the likelihood of infection tends to zero.

We recall some additional standard notation~\cite{lawless1982statistical} that we will use in the remainder of the section. Given some node $j$, infected at
time $t_j$, we define the \emph{survival function} of edge $j\rightarrow i$ as $S(t_i | t_j ; \alpha_{j,i}) = 1-F(t_i | t_j ; \alpha_{j,i})$ where
$F(t_i | t_j ; \alpha_{j,i}) = \int_{t_j}^{t_i} f(t | t_j ; \alpha_{j,i})\,dt$ is the cumulative transmission density function, computed from the transmission likelihood.
Finally, the \emph{hazard function}, or instantaneous infection rate, of edge $j\rightarrow i$ is the ratio $H(t_i | t_j ; \alpha_{j,i}) = f(t_i | t_j ; \alpha_{j,i})/$ $S(t_i | t_j ; \alpha_{j,i})$.
We derive the log survival and hazard functions for the three edge transmission models in Table~\ref{tab:log-likelihood-gradients}.

\xhdr{Likelihood of a cascade} Consider some node $i$ in a directed network. Node $i$ can get infected by any of its parents (i.e., nodes pointing to $i$). Once infected, node $i$ can then also spread the contagion to its children (i.e., nodes $i$ points to). As in the independent cascade model~\cite{kempe03maximizing}, we assume that node gets infected once the \emph{first} parent infects it (i.e., a node can get infected only once). Then, the likelihood of infection of node $i$ at time $t_i$ given a collection of previously infected nodes $(t_1,\ldots,t_N | t_k \leq t_i)$ results from summing over the likelihoods of the mutually disjoint events that each node is the first parent that generated the infection event of our node $i$:
\begin{multline}
	f(t_i | t_1, \ldots, t_N \setminus t_i ; \alphs) = \sum_{j : t_j < t_i} f(t_i | t_j ; \alpha_{j,i}) \times \\
	\prod_{j : j \neq k, t_k < t_i} S(t_i | t_k ; \alpha_{k,i}),
\end{multline}
where $\alphs := \{\alpha_{j,i}\,|\, i,j=1,\ldots,n, i \neq j\}$. If we assume that infections are conditionally independent given the parents of the infected nodes, the likelihood of the
infections in a cascade is:
\begin{multline}
	f(\casc^{\leq T^{c}} ; \alphs) = \prod_{t_i\leq T} \sum_{j : t_j < t_i}
	f(t_i | t_j ; \alpha_{j,i}) \times \\
	\prod_{j : j \neq k,  t_k < t_i} S(t_i | t_k ; \alpha_{k,i}),
\end{multline}
where $\casc^{\leq T^{c}}$ denotes the vector of infected nodes in the cascade up to $T^{c}$. Removing the condition $k \neq j$ makes the product
independent of $j$,
\begin{multline} \label{eq:casc}
	f(\casc^{\leq T^{c}}; \alphs) = \prod_{t_i\leq T} \prod_{k : t_k < t_i}
	S(t_i | t_k ; \alpha_{k,i}) \times \\
	\sum_{j : t_j < t_i} \frac{f(t_i | t_j ; \alpha_{j,i})}{S(t_i | t_j ; \alpha_{j,i})}.
\end{multline}
The fact that some nodes are \emph{not} infected during the observation window is also informative. We therefore add multiplicative survival terms to
Eq.~\ref{eq:casc} and rearrange with hazard functions:
\begin{align} \label{eq:loglikelihood}
	f(\casc;\alphs) &= \prod_{i : t_i \leq T} \prod_{m : t_m > T} S(T | t_i ; \alpha_{i,m}) \times \\ &\prod_{k : t_k < t_i} S(t_i | t_k ; \alpha_{k,i})  \left( \sum_{j : t_j < t_i} H(t_i | t_j ; \alpha_{j,i}) \right).
\end{align}

Perhaps surprisingly, our continuous time model of diffusion is a particular case of Aalen'{}s additive regression model, frequently used in survival theory analysis~\cite{aalen2008survival}.
In Aalen'{}s model, the hazard function, or instantaneous infection rate, of node $i$ is parametrized as $\alpha_{i,0}(t) + \mathbf{\alpha}(t)^T_{i} \mathbf{s}_i(t)$, where
$\mathbf{\alpha}(t)$ is a vector that accounts for the effect of a collection of observable covariates $\mathbf{s}(t)$ and $\alpha_{i,0}(t)$ is a baseline. It is easy to show
that the hazard function of node $i$ at time $t_i$ for the three pairwise transmission models: exponential, power-law and Rayleigh, has the following form:
\begin{equation} \label{eq:aalen}
	H(t_i | t_1, \ldots, t_N \setminus t_i ; \alphs) = \mathbf{\alpha}_i^T \mathbf{s}_i(t_i ; t_1, \ldots, t_N \setminus t_i),
\end{equation}
where $\mathbf{\alpha}_i = (\alpha_{1,i}, \ldots, \alpha_{N, i})$ accounts for the effect of a collection of observable covariates $\mathbf{s}_i(t_i ; t_1, \ldots, t_N \setminus t_i)$, the covariates
depend on the pairwise transmission model (exponential, power-law or Ray\-leigh) and the previously infected nodes, and the baseline is zero.

\xhdr{Dynamic network inference problem}
Given a static network with constant edge transmission rates $\alpha_{j,i}$, the network inference problem reduces to solving a maximum likelihood problem over set of recorded cascades $C$~\cite{manuel11icml}:
\begin{equation}
	\label{eq:opt-problem-static}
	\begin{array}{ll}
		\mbox{maximize$_{\alphs}$} & \sum_{c \in C} \log f(\casc^c;\alphs) \\
		\mbox{subject to} & \alpha_{j,i} \geq 0,\, i,j=1,\ldots,N, i \neq j,
	\end{array}
\end{equation}
where $\alphs := \{\alpha_{j,i}\,|\, i,j=1,\ldots,n, i \neq j\}$ are the edge transmission rates we aim to infer. The edges of the network are those pairs of nodes with transmission rates $\alpha_{j,i} > 0$.

Now we generalize the network inference problem to dynamic networks with edge transmission rates $\alpha_{j,i}(t)$ that may change over time. To this aim, at any given time $t$, we solve a maximum weighted likelihood problem over the set of recorded cascades by time $t$, $C_{t} = \{\casc^1,\ldots,\casc^{|C_{t}|}\}$:
\begin{equation}
	\label{eq:opt-problem}
	\begin{array}{ll}
		\mbox{maximize$_{\alphs(t)}$} & \sum_{c \in C_t} {w_c(t)} \log f(\casc^c;\alphs(t)) \\
		\mbox{subject to} & \alpha_{j,i}(t) \geq 0,\, i,j=1,\ldots,N, i \neq j,
	\end{array}
\end{equation}
where $w_c(t) \geq 0$ is a weight that penalizes the importance of cascade $c$ based on how old it is at time $t$ and $\alphs(t) := \{\alpha_{j,i}(t)\,|\, i,j=1,\ldots,n, i \neq j\}$ are the variables. The intuition here is that diffusion network smoothly changes over time and that recent cascades have higher importance in determining current network structure than old cascades. Thus, at any point in time, we can solve the above optimization problem to obtain the structure of the diffusion network at that particular time. Next, we show how to efficiently solve the above optimization problem for all time points $t$.

\section{The InfoPath algorithm}
\label{sec:proposed}
% !TEX root = dynamic-network-inference.tex
The problem defined by equation Eq.~\ref{eq:opt-problem} is convex for the three transmission models we consider. Therefore we can aim to find the unique optimal solution at any given time point $t$:
\begin{theorem}[\cite{manuel11icml}]
	Given log-concave survival functions and concave hazard functions in the parameter(s) of the pairwise transmission likelihoods, the network
	inference problem defined by equation Eq.~\ref{eq:opt-problem} is convex in $\alphs$.
\end{theorem}
\begin{algorithm}[t]
\caption{\infopath: the dynamic network inference algorithm\label{alg:dynamic-netrate}}
\begin{algorithmic}
\REQUIRE $C, t, K, T, \rho$
\WHILE{$k < K$}
  \STATE $c_k \leftarrow$ cascade-sampling$(C, t, T)$;
  \FORALL{$(j,i) : t_j^{c_k} < t_i^{c_k}$}
    \STATE $\alpha_{j,i}^k = \left( \alpha_{j, i}^{k-1} - \gamma_{k} \nabla_{\alpha_{j,i}} L_{c_k}(\alphs^{k-1}) \right)^{+}$;
  \ENDFOR
  \FORALL{$(j,i) : \alpha_{j,i}^{k-1} > 0, t_j^{c_k} \rightarrow \infty$}
    \STATE $\alpha_{j,i}^k = \rho \alpha_{j, i}^{k-1}$;
  \ENDFOR
  \STATE k = k+1;
  \ENDWHILE
\STATE $\alphs^{*} \leftarrow \alphs^{K-1}$;
\RETURN $\alphs^{*}$;
\end{algorithmic}
\end{algorithm}

Stochastic gradient (SG) methods have been shown to be extremely successful for taking advantage of the structure exhibited by the optimization problem stated in Eq.~\ref{eq:opt-problem}. They have received increasing attention in the machine learning literature~\cite{agarwal2011distributed, bach2011non,blatt2008convergent, duchi2011ergodic,roux2012stochastic}. Although many optimization methods based on stochastic gradient descent have been proposed,
we have found that in practice the basic projected stochastic gradient method~\cite{robbins1951stochastic} works well for our problem. Other
more sophisticated methods, like the stochastic average gradient~\cite{roux2012stochastic} or incremental average gradient~\cite{blatt2008convergent} do
not offer a significant advantage. Therefore, we proceed with the basic stochastic gradient method in the remainder of the paper.
\begin{figure*}[t]
	\centering
	\subfigure[Slab]{\includegraphics[width=0.23\textwidth]{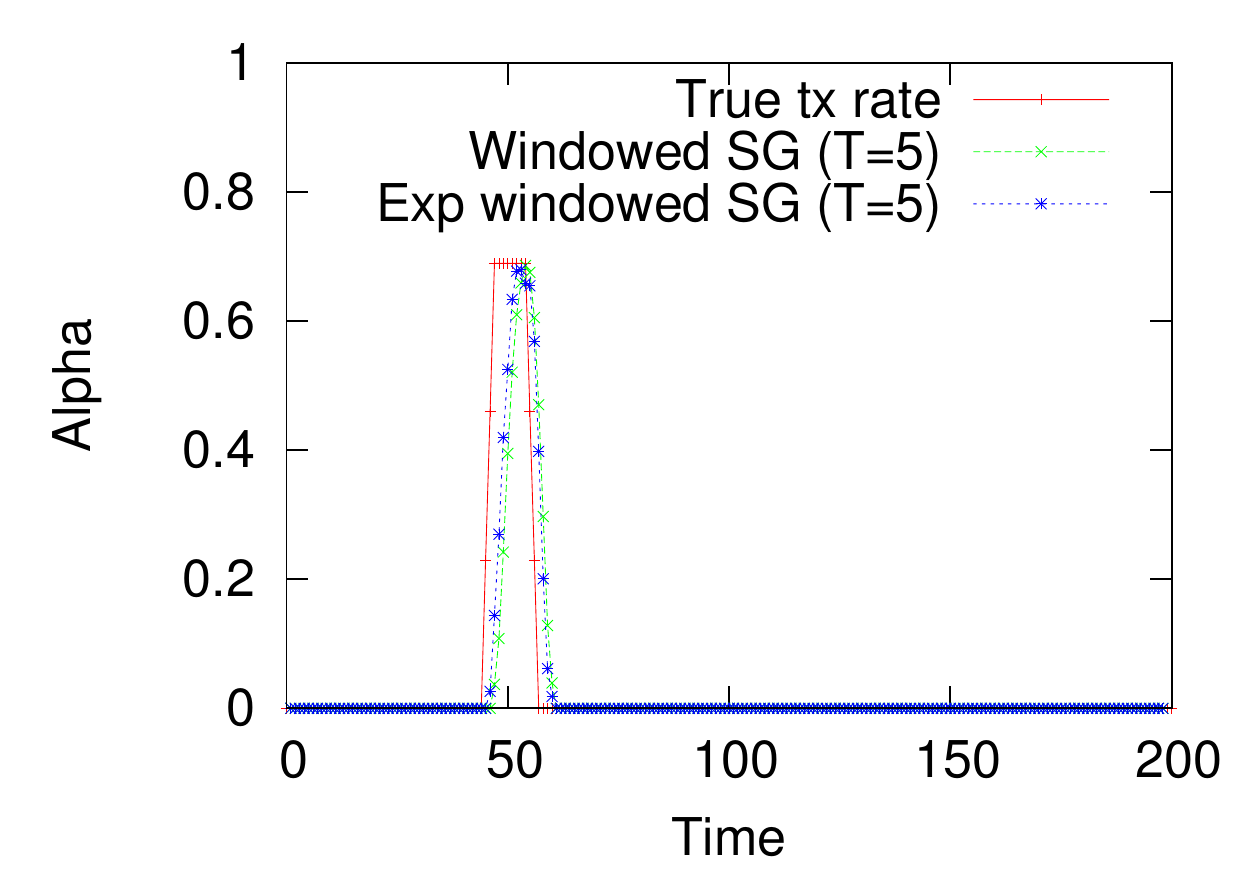}
	\label{fig:slab-trend-alpha}}
	\subfigure[Square]{\includegraphics[width=0.23\textwidth]{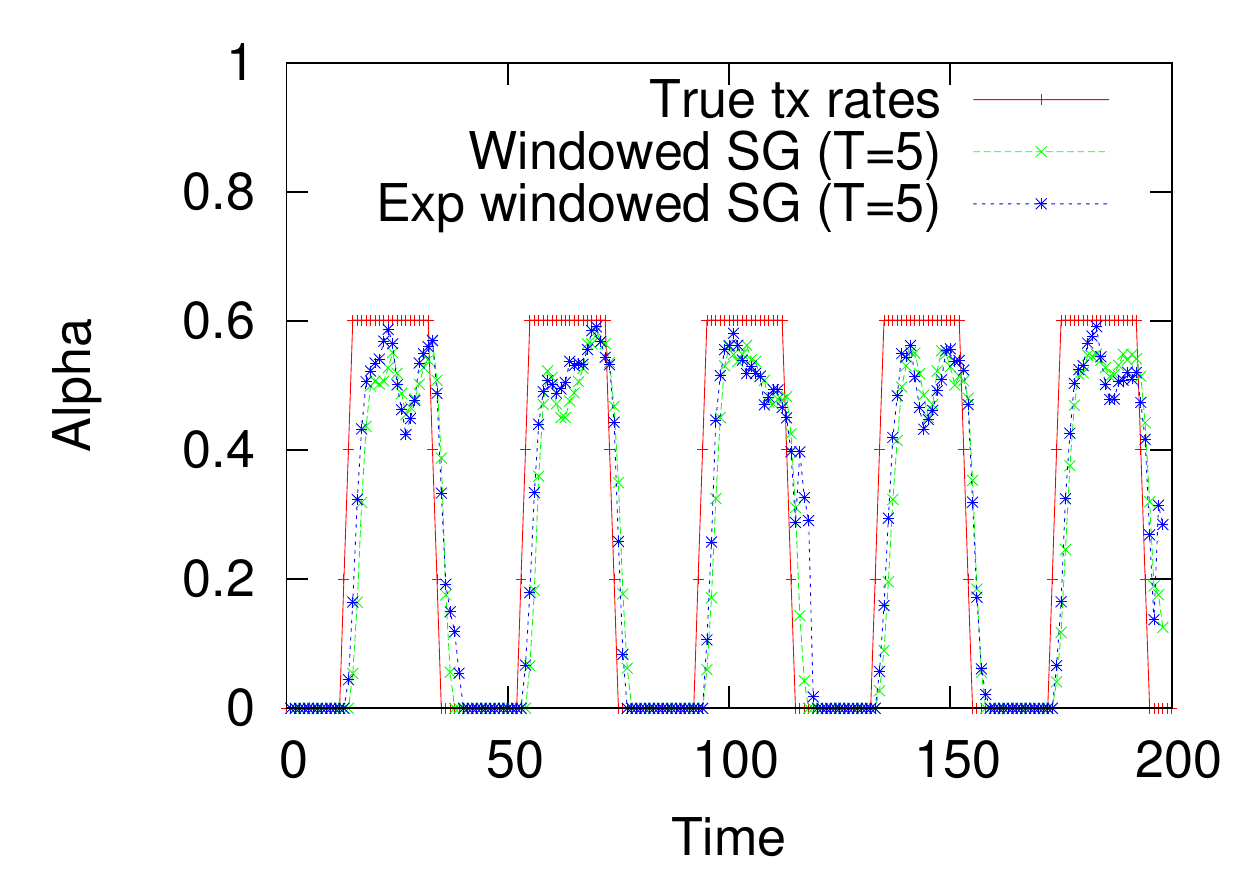}
	\label{fig:square-trend-alpha}}
	 \subfigure[Chainsaw]{\includegraphics[width=0.23\textwidth]{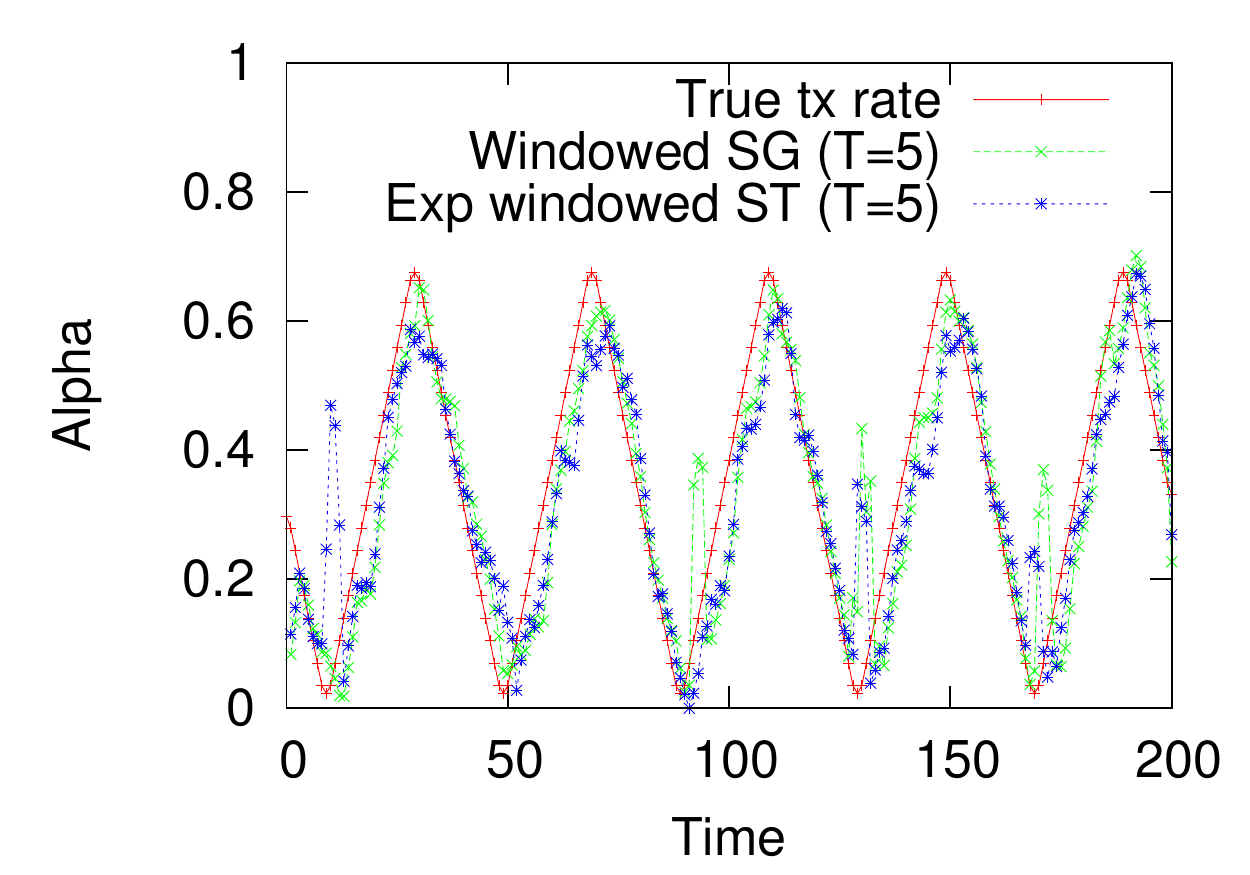}
	\label{fig:chainsaw-trend-alpha}}
	\subfigure[Hump]{\includegraphics[width=0.23\textwidth]{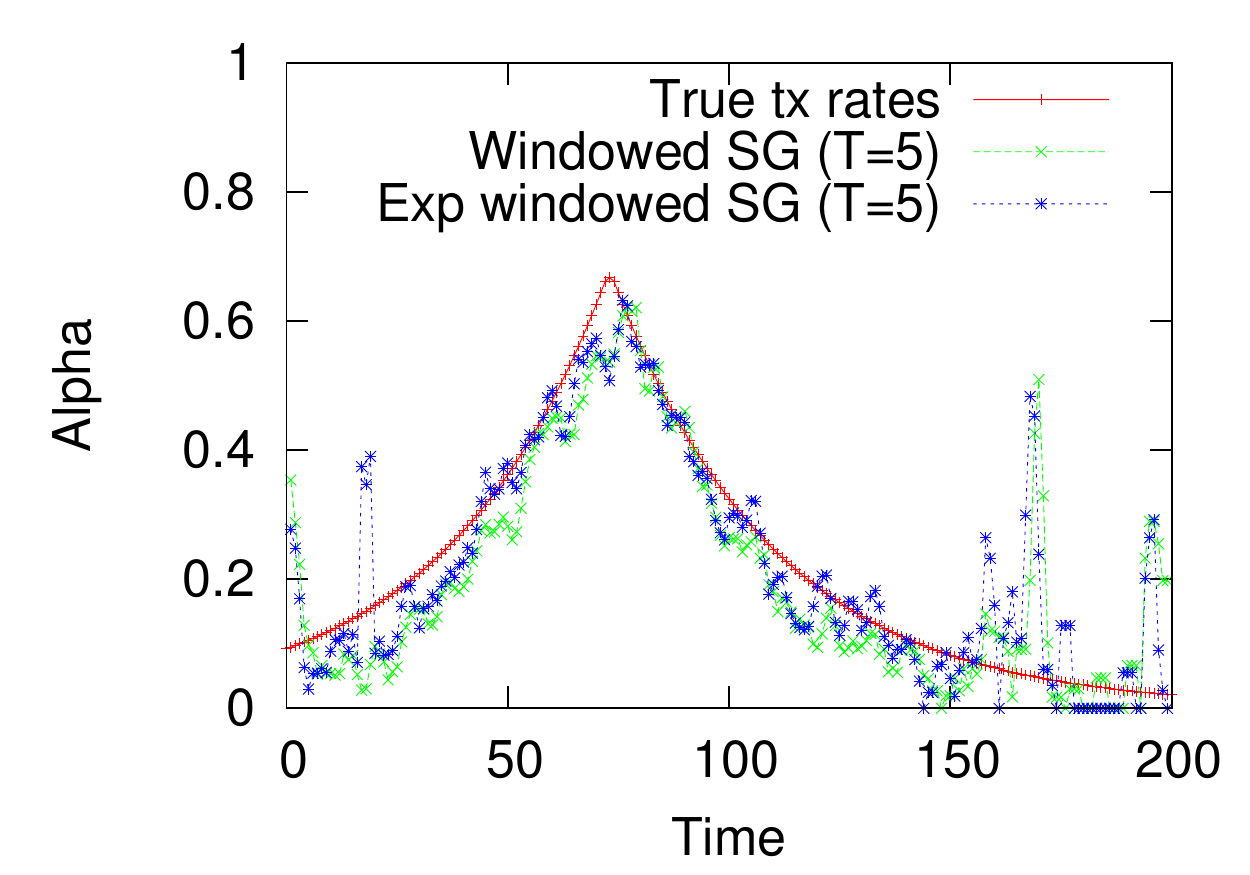}
	\label{fig:exp-trend-alpha}}
	\vspace{-3mm}
 	\caption{True and inferred edge transmission rates for edges with different 4 transmission rate evolution patterns: (a) Slab, (b) Square, (c) Chainsaw, (d) Hump.
 	Results are for the Kronecker core-periphery with exponential edge transmission model for 200 time units with 1,000 cascades per time unit. Our \infopath method is able to track the evolving edge transmission rates over time. \infopath works better for continuously evolving edge transmission rates (c, d).} \label{fig:trends-alphas-estimation}
	\vspace{-3mm}
\end{figure*}

\xhdr{Projected Stochastic Gradient} The projected stochastic gradient method~\cite{robbins1951stochastic} uses iterations of the form:
\begin{equation} \label{eq:update}
\alpha_{j, i}^{k}(t) = \left( \alpha_{j, i}^{k-1}(t) - \gamma_{k} \nabla_{\alpha_{j,i}} L_{c_k}(\alphs^{k-1}(t)) \right)^{+},
\end{equation}
where $\nabla_{\alpha_{j,i}} L_{c_k}(\cdot)$ is the gradient of the log-likelihood $L_{c}(\cdot)$ with respect to the edge transmission rate $\alpha_{j,i}$, $\gamma_k$ is a
step-size, $(z)^{+} = \max(0, z)$, and cascade $c_k$ is sampled (with replacement) from $C_{t}$. The gradients for all three edge transmission models are given in Table~\ref{tab:log-likelihood-gradients}.

Note that instead of using all historic data and then explicitly penalizing each cascade by a different weighting factor $w_c(t)$, we use a different, more scalable approach. We sample cascades with replacement, where the probability of a cascade being sampled decays with the age of the cascade. This way recent cascades get sampled more often and thus implicitly hold higher importance when inferring the network. In practice, we achieve a significant speed up by using this approach.
Moreover, in our dynamic network in\-fe\-rence problem, the edge transmission rates usually vary smoothly. This means that stochastic gradient descent is a perfect method as we can use the inferred network from the previous time step as initialization for the inference procedure in the current time  step. We find that setting the starting point $\alpha_{j,i}^{0}$ of each edge transmission rate $\alpha_{j,i}$ to the last outputted estimate of the transmission rate \-allows us to further speed up the algorithm.

Importantly, in each iteration $k$ of the projected stochastic gra\-dient method, we only need to compute the gradients $\nabla_{\alpha_{j,i}} L_{c_k}(\alphs^{k})$ for edges
$(j, i)$ such that node $j$ has been infected in cascade $c_k$, and the iteration cost and convergence rate are independent of $|C|$~\cite{bach2011non, nemirovski2009robust}.
Rigorous theoretical analysis of convergence turns out to be a challenging problem which we leave for future work.
%A rigorous theoretical analysis of convergence can be found elsewhere~\cite{robbins1951stochastic}, and it is beyond the scope of this paper. 
However, we point out that standard analyses~\cite{robbins1951stochastic} typically assume the gradients $\nabla_{\alphs} L_{c}(\alphs^{k})$ to be either bounded above by a constant $M$, where $||\nabla_{\alphs}
L_{c}(\alphs)|| \leq M$, or Lipschitz-continuous with constant $L$, $||\nabla_{\alphs} L_{c}(\alphs_{2}) - \nabla_{\alphs} L_{c}(\alphs_{1})|| \leq L || \alphs_{2} - \alphs_{1} ||$.
In our case, these conditions are violated if at any iteration $k$, there is a node $i$ infected in cascade $c_k$ such that $H(t^{c_k}_i | t^{c_k}_j ; \alpha^{k-1}_{j,i}) =
0\,\,\, \forall j : t^{c_k}_j < t^{c_k}_i$, \ie, node $i$ has no parents that \emph{explain} the infection at $t^{c_k}_i$, and the objective function is positively unbounded. In practice, we obtain a good performance and avoid such scenario by bounding below each \emph{feasible} transmission rate, $\alpha_{j,i} \geq \varepsilon$. An edge transmission rate $\alpha_{j,i}$ is \emph{feasible} if there 
is at least one cascade in which both node $j$ and $i$ get infected. When outputting the final solution, we simply omit edges with transmission rates $\varepsilon$.
\begin{figure}[t]
	\centering
	\subfigure[P-R (C-P, \expo)]{\includegraphics[width=0.22\textwidth]{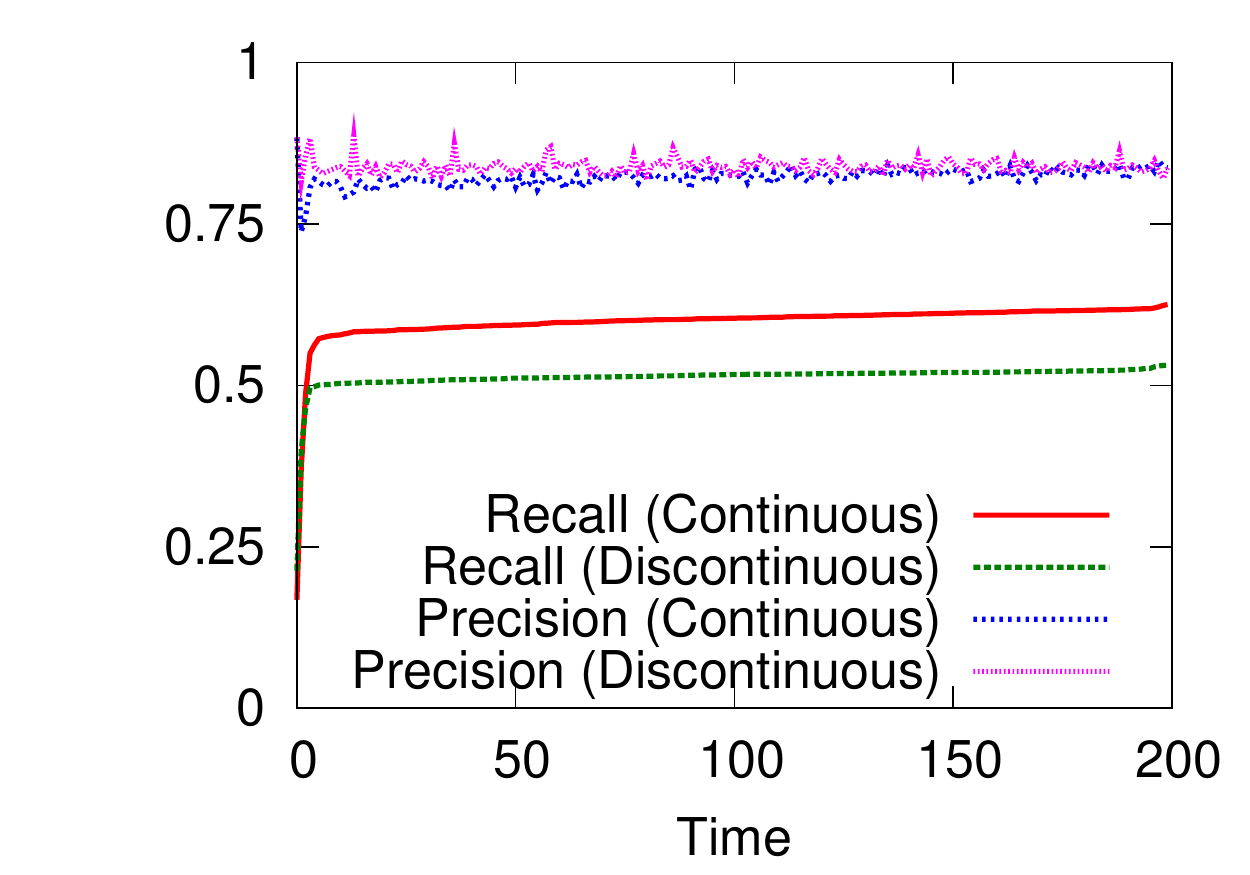}
	\label{fig:pr-vs-running-time}}
	\subfigure[P-R (HI, \ray)]{\includegraphics[width=0.22\textwidth]{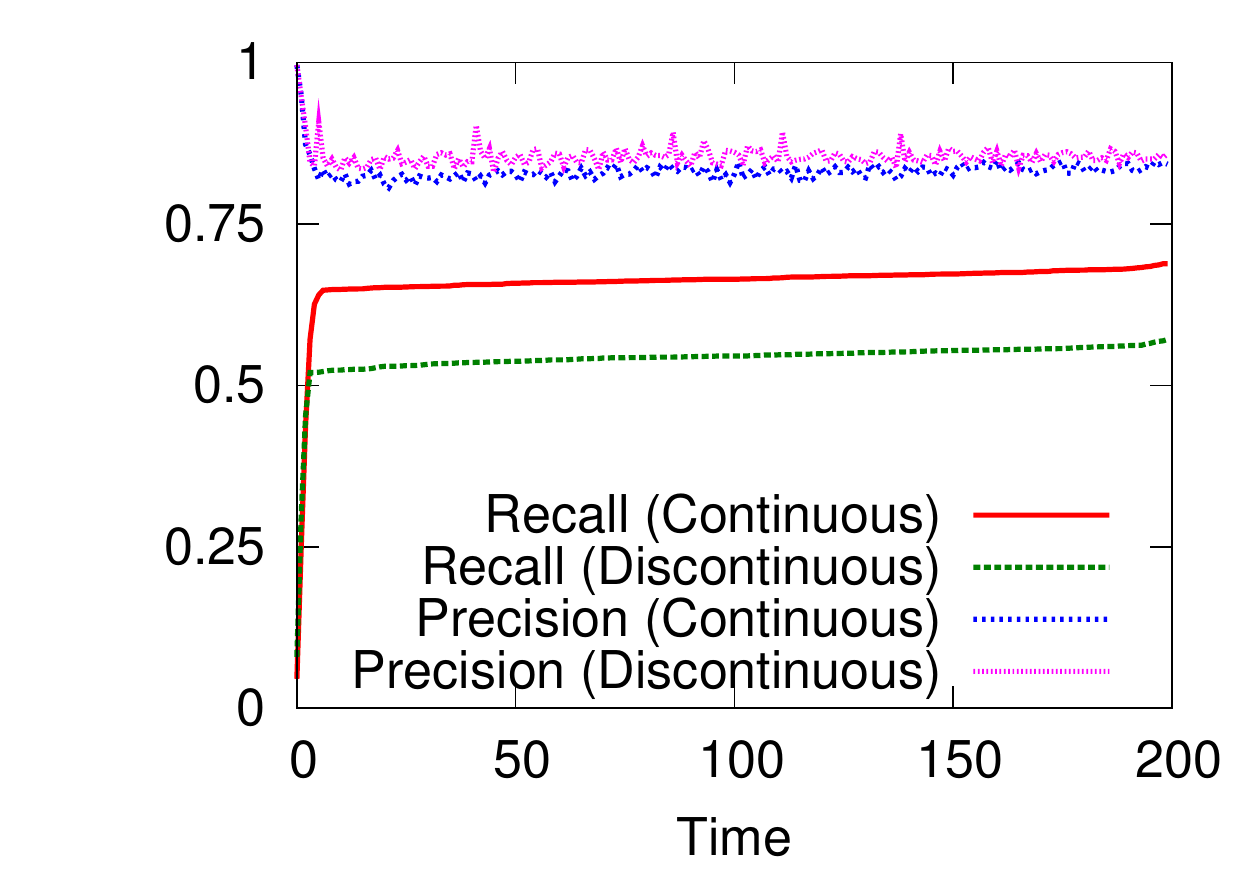}
	\label{fig:pr-vs-running-time}}
	\\
	\subfigure[Accuracy (C-P, \expo)]{\includegraphics[width=0.22\textwidth]{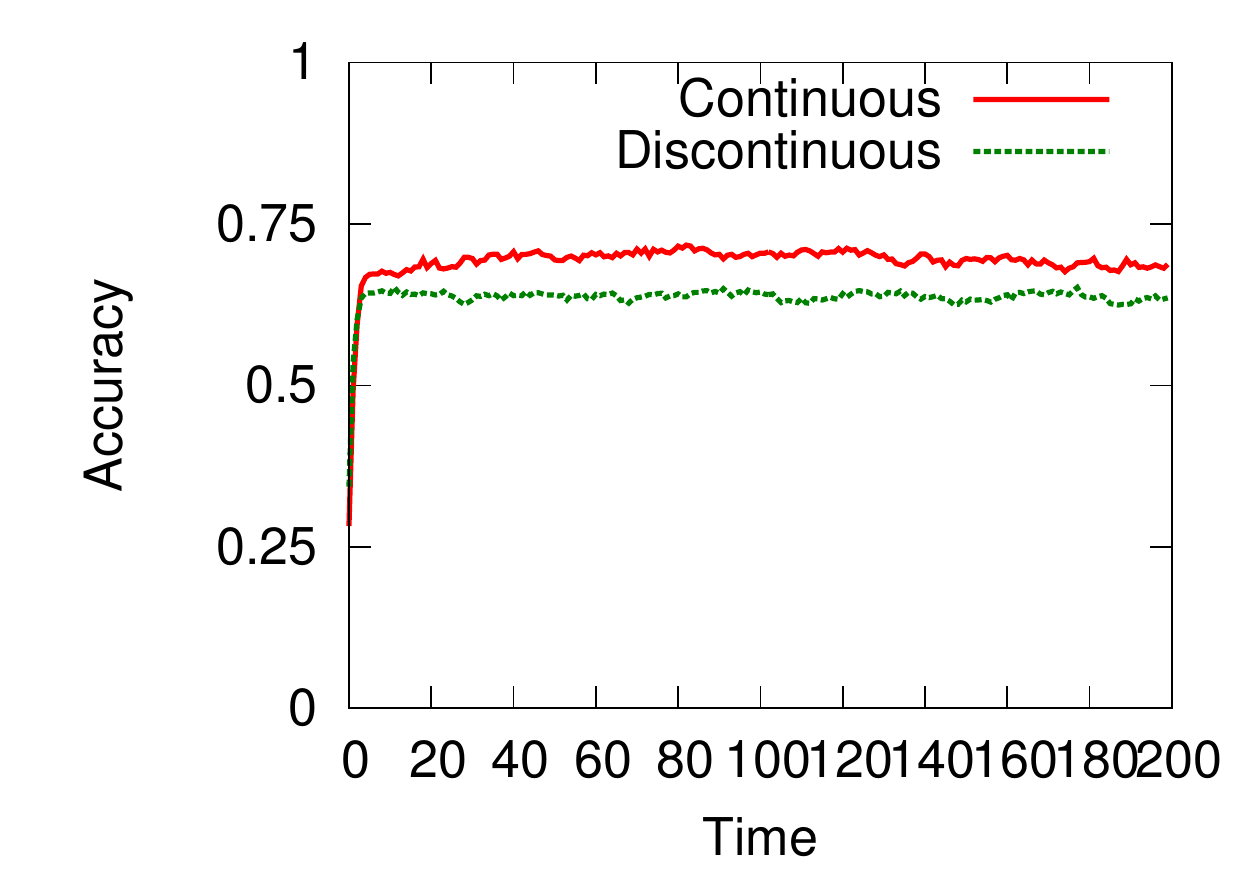}
	\label{fig:acc-vs-running-time}}
	\subfigure[Accuracy (HI, \ray)]{\includegraphics[width=0.22\textwidth]{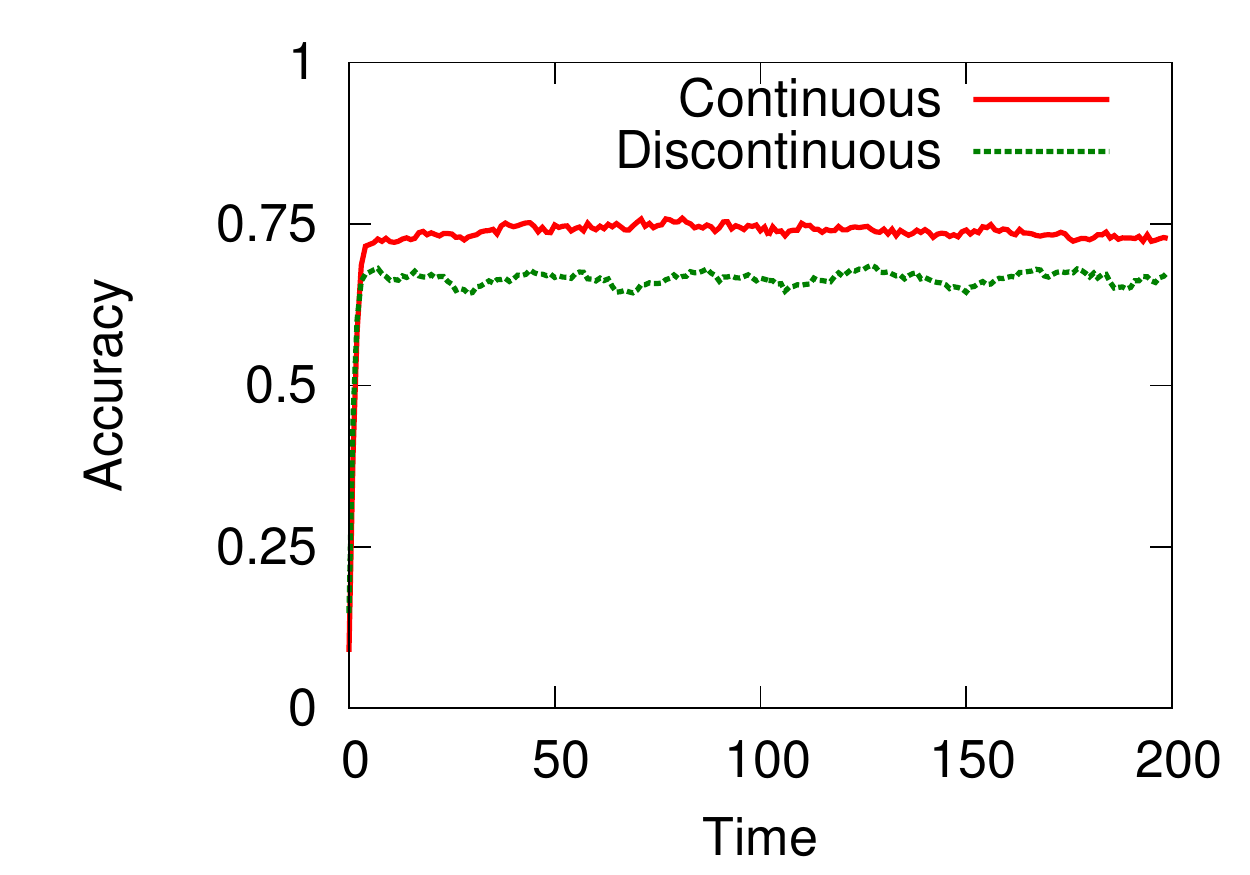}
	\label{fig:acc-vs-running-time}}
	\\
	\subfigure[MSE (C-P, \expo)]{\includegraphics[width=0.22\textwidth]{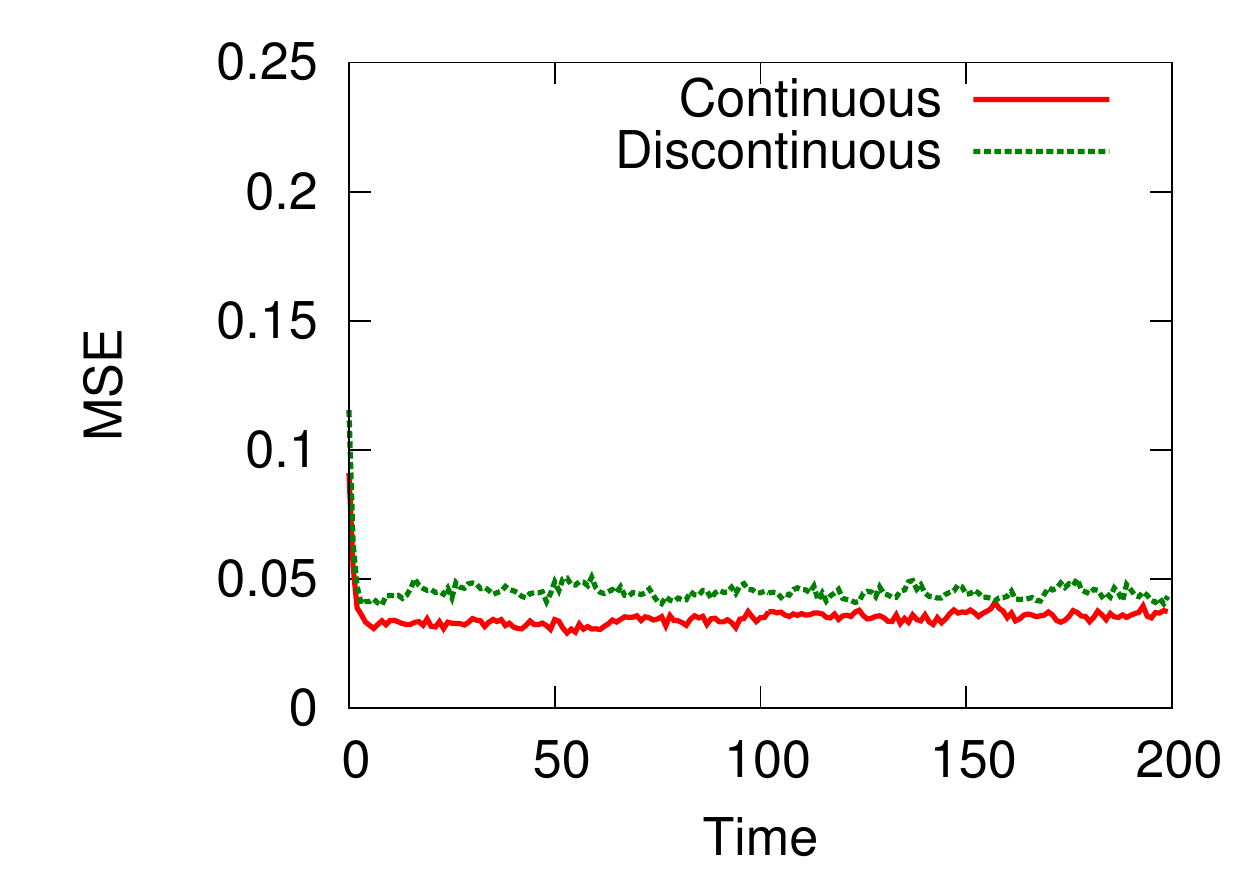}
	\label{fig:mse-vs-running-time}}
	\subfigure[MSE (HI., \ray)]{\includegraphics[width=0.22\textwidth]{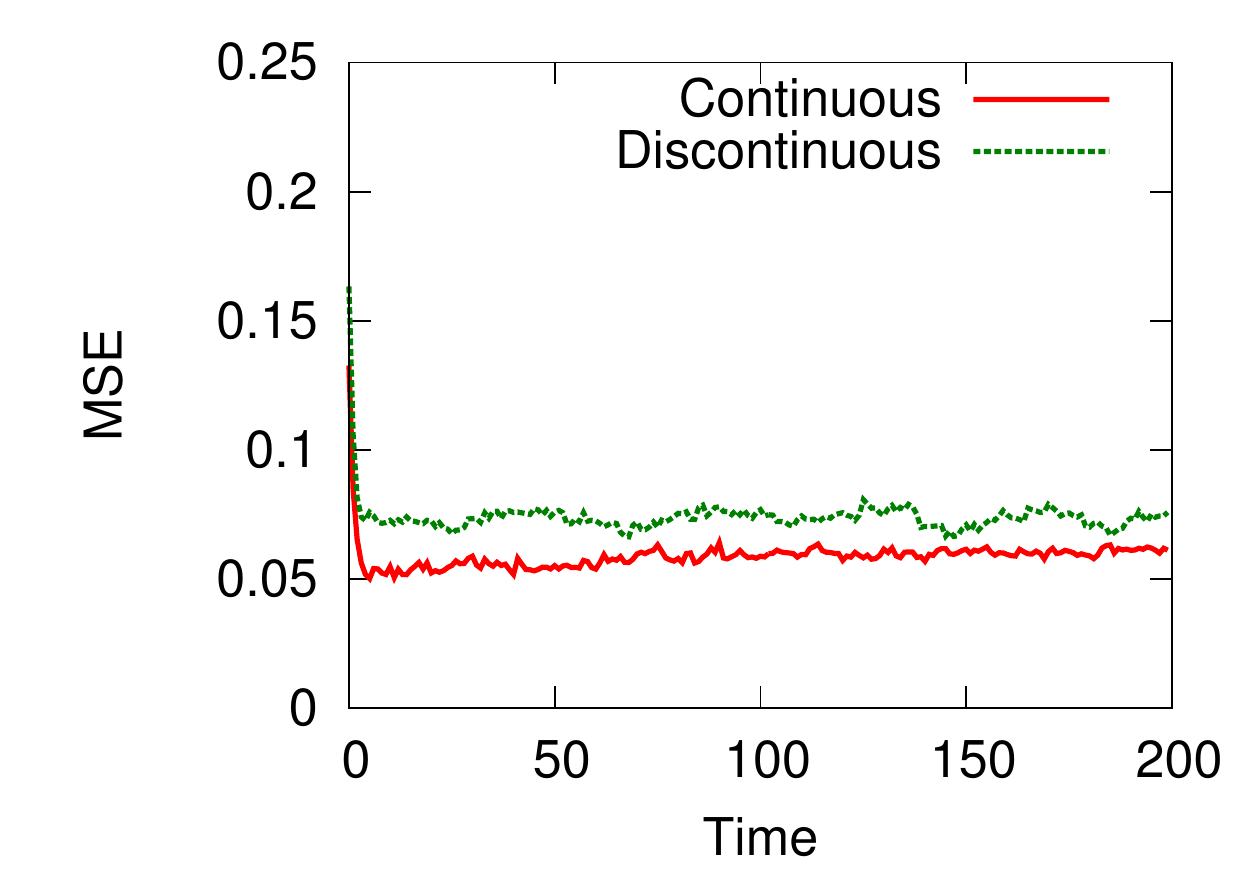}
	\label{fig:mse-vs-running-time}}
	\vspace{-3mm}
 	\caption{Precision and Recall (P-R), Accuracy and Mean Squared Error (MSE) of our \infopath method against time. (a,c,e): Core-periphery (C-P)
	Kronecker network with exponential edge transmission model (b, d, f), and Hierarchical (HI) Kronecker network with Rayleigh  edge transmission model. Performance on Type I (Chainsaw, Hump) and Type II (Slab, Square) edge transmission rate evolution patterns is plotted.%, and 1,000 cascades per unit time were recorded.
	} \label{fig:performance-vs-time}
	\vspace{-3mm} 	
\end{figure}

\xhdr{Aging edges} Suppose we solve the dynamic network inference problem for a given time $t$ using the projected SG method. In each iteration $k$, we only
update edge transmission rate $\alpha_{j,i}^{k}$ if node $j$ has been infected in cascade $c_k$.
Now, suppose a few edge transmission rates $\alpha_{j, i}^{0}$ are greater than zero for a given node $j$, \ie, their last outputted estimates before $t$ have been positive. Then, if
node $j$ turns to be infected in many sampled cascades at time $t$ but it never transmits information to any of its neighbors $i$ in the future, the edge transmission rates $\alpha_{j,i}^{k}$ will eventually converge to zero for large $k$.
However, if node $j$ is never infected for any of the future cascades, then none of edge transmission rates $\alpha_{j,i}$ will be updated, and they will remain positive. So, if node $j$ is never infected in subsequent future cascades, the transmission rates $\alpha_{j,i}$ will remain positive forever.
However, we would like these \emph{unused} edges $(j, i)$ to decay and eventually vanish, or equivalently the transmission rates $\alpha_{j,i}$ to converge to zero. To achieve this, we multiply transmission rates of \emph{unused} edges by \emph{aging} factor $\rho$ every time we solve the dynamic network inference problem. We use
$\rho = 0.95$ in all experiments.

\xhdr{Cascade sampling} In Eq.~\ref{eq:update}, instead of sampling cascades uniformly at random and explicitly penalizing each cascade by a different weighting factor $w_c(t)$, we achieve a significant speed up by sampling cascades using a sampling procedure that penalizes old cascades and considers $w_c(t)=1$ for all cascades.
There are many possible sampling schemes. In practice, and for simplicity, we use windowed uniform sampling or windowed exponential sampling. Windowed means that when solving the network inference problem for time $t$, we only sample (uniformly or exponentially) cascades that have started in a sampling time window $(t-T, T)$.
Here, we find an an important tradeoff. The shorter is the sampling time window $T$ in the stochastic gradient descend, the quicker our algorithm is tracking changes in the
edge transmission rates. However, short sampling time window results in less reliable estimates because of sampling from a smaller universe of distinct cascades. Therefore, in order to be able to track changes quickly, we would need to observe many cascades over time.

\section{Experimental evaluation}
\label{sec:evaluation}
% !TEX root = dynamic-network-inference.tex
We evaluate the performance of \infopath on time-varying synthetic networks that mimic the structure of real networks as well as on a dataset of more than 179 million information cascades extracted from 300 million blogs and news articles from 3.3 million media sites over a period of one year, from March 2011 till February 2012. All the data, code and additional results are available at the supporting website~\cite{website13}. 

\subsection{Experiments on synthetic data} \label{sec:synthetic-experiments}
The goal of the experiments with synthetic data is to understand how temporal changes in a network affect the performance of our algorithm. We aim to detect not only
when an edge appears (\ie, its transmission rate becomes $> 0$) or disappears (\ie, its transmission rate becomes $0$) but also provide instantaneous transmission rate estimates that track the true edge transmission rates over time.

\xhdr{Experimental setup} First, we generate synthetic networks using Kronecker graph models of directed real-world networks~\cite{leskovec2010kronecker}. For all our experiments, we consider two different Kronecker networks, both with 1,024 nodes and 2,048 edges: A {\em core-periphery} Kronecker network with parameter matrix $[0.9, 0.5; 0.5, 0.3]$) and a {\em hierarchical} Kronecker network with parameters $[0.9, 0.1; 0.1, 0.9]$.

The next step is to make each edge to follow a particular edge transmission rate evolution pattern. Our goal later will be to recover the network as well as the evolution of the transmission rate 
of each individual edge.

%
%We continue by assigning transmission rate trends to the edges of each network $G^*$ to make them time-varying. 
%
%We assign each edge in the network an evolution pattern chosen uniformly at random from the set of 5 patterns. 
%
We consider five edge evolution patterns: Slab, Square, Chainsaw, Hump and constant (see Figure~\ref{fig:trends-alphas-estimation}). Slab and Hump patterns model outgoing connections of sites that become popular for a short period of time. Square and Chainsaw patterns model incoming connections to sites that perform updates periodically at specific times of the day or days of the week. Constant pattern represents connections between sites that interact at any time and during a long period of time, usually large media sites. 
We consider Chainsaw, Hump and Continuous to be examples of {\em Type I} pattern, without discontinuities, and Slab and Square to be examples of Type II pattern, with discontinuities. 

We assign to each edge in the network an evolution pattern chosen uniformly at random from the set of the above 5 patterns. Then, we generate transmission rate values $\alpha^*_{j,i}(t)$ for 
each edge according to its chosen evolution pattern. The evolving edge transmission rate $\alpha^*_{j,i}(t)$ models how quickly information spreads from one node to another. Finally, we 
generate 1,000 information cascades per time step. For each cascade we randomly pick the cascade initiator node.
%we select root nodes uniformly at random and generate and record cascades over each network $G^*$.

Given the node infection times from the recorded cascades, our goal then is to find the true edges of the network and for each edge discover its transmission rate evolution pattern. In other words,
inferring how each edge transmission rate $\alpha(t)$ evolves over time. Figure~\ref{fig:trends-alphas-estimation} shows the true and inferred edge transmission rates for four different edges, each with a different evolution pattern: Slab, Square, Chainsaw and Hump. 
%, in a 512 node, 1,024 edge core-periphery Kronecker network (parameter matrix $[0.9, 0.5; 0.5, 0.3]$) \cite{jure08ncp} with 20\% of the edges following each of the five rate trends.
%We generated and recorded an average of 1,000 cascades per time unit using an exponential pairwise transmission model.
%
Observe that \infopath is able to track the evolving edge transmission rates over time for all evolution patterns. \infopath gives near perfect performance when edge transmission rate evolves continuously (Chainsaw, Hump). Interestingly, even when the edge transmission rate evolves dis\-con\-tinuous\-ly (Slab, Square), \infopath manages to track it.

\xhdr{Accuracy of \infopath} We evaluate the \infopath method quantitatively by computing four different measures for every time step: Precision, Recall and Accuracy of inferred edges as well as the Mean Squared Error (MSE) in the edge transmission rate. Precision at time $t$ is the fraction of edges in the inferred network $\hat{G}(t)$ present in the true network $G^*(t)$. Recall at time $t$ is the fraction of edges of the true network $G^*(t)$ present in the inferred network $\hat{G}(t)$. And accuracy at time $t$ is defined as
$$1-\frac{\sum_{i,j} |I(\alpha^*_{i,j}(t))-I(\hat{\alpha}_{i,j}(t))|}{\sum_{i,j}I(\alpha^*_{i,j}(t)) + I(\hat{\alpha}_{i,j}(t))},$$
where $\alpha^*(t)$ is the true transmission rate at time $t$, $\hat{\alpha}(t)$ is the estimated transmission rate at time $t$, and $I(\alpha(t))=1$ if $\alpha(t) > 0$ and $I(\alpha(t))=0$ otherwise. Inferred networks with no edges or only false edges would have zero accuracy. Last, Mean Squared Error (MSE) at time $t$ is defined as $E\big[||\alpha^*(t)-\hat{\alpha}(t)||^{2}\big]$, where $\alpha^*(t)$ is the true edge transmission rate at time $t$ and $\hat{\alpha}(t)$ is the estimated transmission rate.

% XXXXX
Figure~\ref{fig:performance-vs-time} shows Precision, Recall, Accuracy, and MSE over time for the time-varying core-periphery Kronecker network with exponential edge transmission model, and hierarchical Kronecker network with Rayleigh edge transmission model.
%We generated continuous (chainsaw, exponential) and discontinuous (slab, square) trends for transmission rates, $\alpha^*_{j,i}(t) \in [0, 1]$ for all $t$, and we recorded 1,000 cascades per unit time. 
Observe that the performance of our method is stable across time, and as mentioned before, continuous evolution patterns are easier to track and estimate than discontinuous ones.

\xhdr{Accuracy vs. running time in static networks} Our stochastic gradient descend based method, \infopath, can be also used to speed-up inference of static networks. In
such scenario, stochastic gradient descend processes cascades in a random round-robin fashion. 
%cascades are sampled uniformly at random in the stochastic gradient descend. 
Here, we compare \infopath to the state of the art methods for inference of static networks: \netinf~\cite{manuel10netinf} and \netrate~\cite{manuel11icml}. First, we compare the methods by computing the accuracy against running time.
Second, we compare \infopath to \netrate in terms of mean squared error of the estimated transmission rates against the running time. We omit \netinf from this last comparison since it only infers the network structure (and no edge transmission rates). For the sake of fairness in the running time comparison we implemented all methods in C++. Our C++ implementation of \netrate is much faster than the public Matlab implementation.
%and since we implemented our method in C++, as the public
%domain implementation of \netinf, we developed our own full gradient (non-stochastic) descent implementation of \netrate in C++, faster than the public code which was developed in Matlab.
%
\begin{figure}[t]
	\centering
	 \subfigure[Accuracy]{\includegraphics[width=0.22\textwidth]{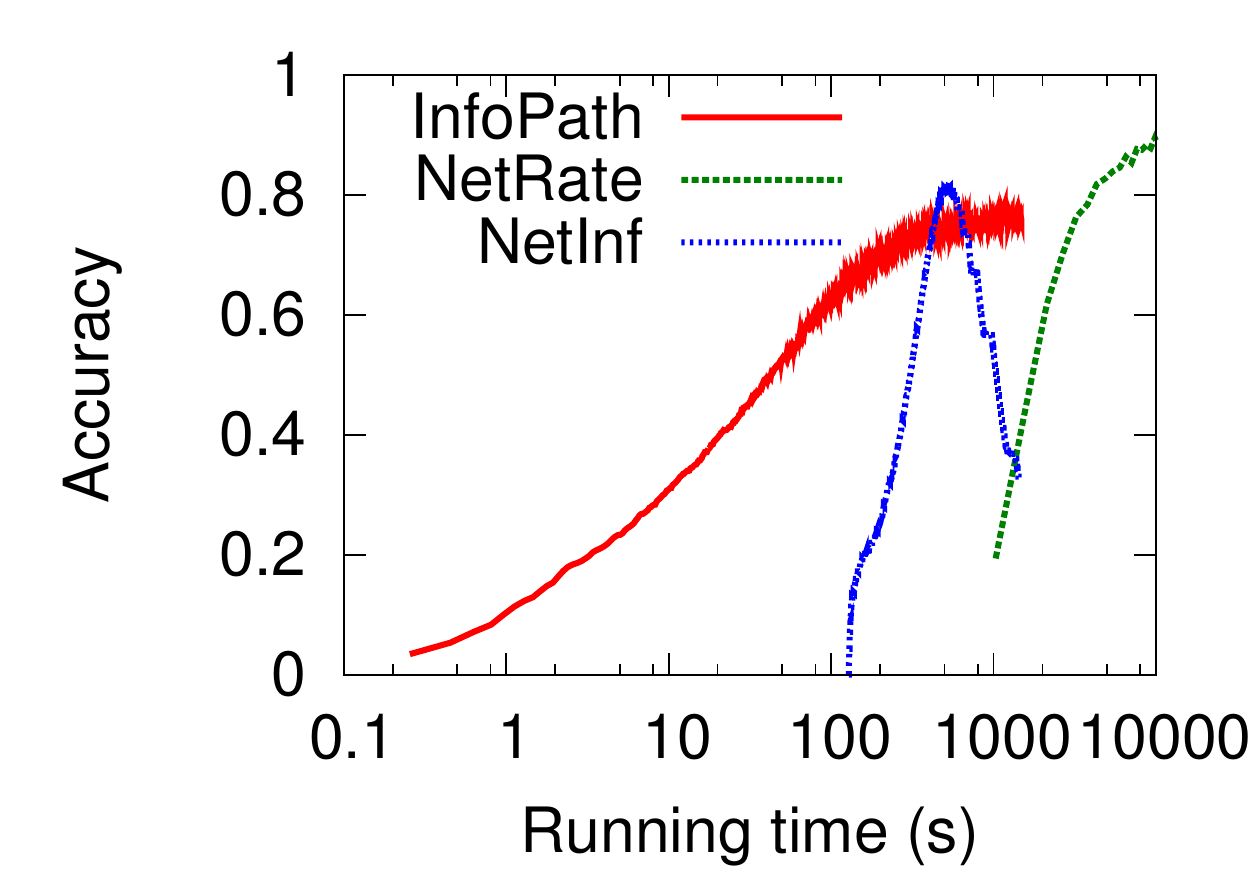}
	\label{fig:acc-vs-running-time}}
	 \subfigure[MSE]{\includegraphics[width=0.22\textwidth]{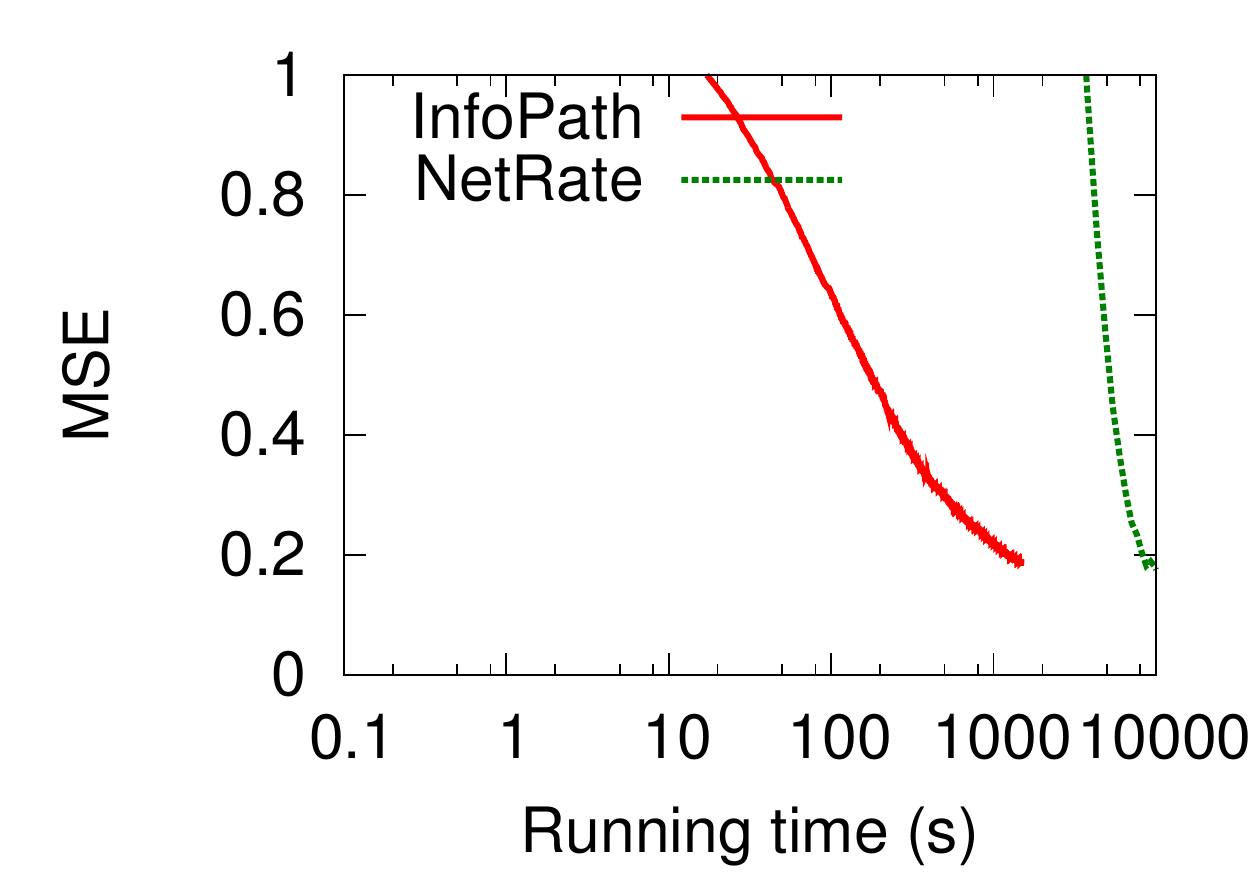}
	\label{fig:mse-vs-running-time}}
	\vspace{-3mm}
	\caption{Accuracy and Mean Squared Error (MSE) against running time for a 1,024 node, 2,048 edge time-invariant core-periphery
	Kronecker network with power-law edge transmission model and 5,000 cascades. Longer running times mean the algorithms run for more iterations. \infopath and \netrate improve accuracy until convergence. However, \infopath achieves the same level of performance 10-100 times faster. 
	} \label{fig:performance-vs-running-time}
	\vspace{-3mm}
\end{figure}

Figure~\ref{fig:performance-vs-running-time} compares Accuracy and MSE against running time for the three network inference methods for a static core-periphery Kronecker network.
% for a core-periphery network with 1,024 nodes and 2,048 edges, power-law model
%($\alpha \in (0,2)$) and 5,000 cascades. Longer running times correspond to more iterations (in our method and \netrate) or higher $k$ (number of inferred edges in \netinf).
%We run \ only once, sampling cascades uniformly at random. 
\infopath is about 10 to 100 times faster than \netrate and as fast as \netinf, while achieving the same accuracy as \netrate. Importantly, \infopath and \netrate always improve accuracy with the running time, until convergence. In terms of MSE, \infopath achieves lower MSE values much quicker than \netrate.

\subsection{Experiments on real data}

The emergence of specific information pathways often depends on the information content of the news that is propagating~\cite{seth2012kdd,romero11twitter}. For example, a real world event may occur for a limited period of time and thus news related to the event spread quicker and to larger parts of the network around such time period.
At any given time, there are many different real world events, topics, and content that propagates through the Web, leading to different emerging and vanishing information pathways, and thus an underlying time-varying network.
In order to better understand these temporal changes, we aim to reconstruct time-varying networks and the information pathways for particular real world events and topics.

\xhdr{Dataset description} We experiment with more than 300 million blog posts and news articles collected from 3.3 million websites over a period of one year, from March 2011 till February 2012.
We trace the flow of information using \emph{memes}~\cite{leskovec2009kdd}. \emph{Memes} are a short textual phrases (like, ``lipstick on a pig'') that travel through the Web.
We consider each meme $m$ as a separate information cascade $c_{m}$. Since all documents which contain memes are time-stamped, a cascade $c_{m}$ is simply a record of times when sites first mentioned meme $m$. We extracted more than 179 million memes, longer than four words. Out of these, 34 million distinct memes appeared at least twice, resulting in 34 million different information cascades.

\xhdr{Experimental setup}
Our aim is to consider sites that actively spread memes over the Web. We achieve this by selecting top 5,000 sites in terms of the number of memes they mentioned. Moreover, we are interested in inferring dynamic networks related to particular topics or events. So, we assume we are also given a keyword query $Q$ related to the event/topic of interest. When inferring a network for a given query $Q$, we only consider documents (and the memes they mention) that include keywords $Q$.
%filter out any document (or post) of those 5,000 sites that does
%not contain a particular keyword or set of keywords that are representative of a given topic or real world event. 
%Then, we build meme cascades using only the memes in those posts.
Then, we build information cascades using only those memes and apply the \infopath algorithm to infer the edges and evolving edge transmission rates. The edge transmission rates explain the propagation of information related to a given topic or real world event $Q$. For each query $Q$ we infer one network per day. Table~\ref{tab:cascades} summarizes the number of sites and meme cascades for several topics
and real world events.\footnote{Additional time-varying diffusion networks for other topics and news events are available at the supporting website~\cite{website13}}.

%\footnote{Additional plots for other topics and news events, and time-varying diffusion networks at a daily resolution are available in GEXF format at \url{http://snap.stanford.edu/proj/dynamic/}}.

\xhdr{Implementation and scalability} We developed an efficient distributed implementation of our \infopath algorithm in C++ based on the network analysis library SNAP~\cite{snap}. We deployed the implementation in a cluster with 1,000 CPU cores and 6 TB of RAM.
With this setup, we considered 38 different topics/events $Q$. For each topic, we inferred a time-varying network with a daily temporal resolution for a period of one year, from March 2011 to February 2012. Each network has thousands of nodes and is based on hundreds of thousands of cascades.  Inferring 38 different time-varying networks took less than 4 hours on our cluster. Note that this is equivalent to solving Eq.~\ref{eq:opt-problem} more than 13,000 times (38 x 365) for millions of pairwise transmission rates. 
We also tested our algorithm on larger datasets. For example, for ``Occupy Wall Street movement'', we were able to infer a 43,415-node time-varying network over a period of 18 months, from January 2011 to June 2012, using 1,381,793 information cascades. % JURE: "in less than 5 hours" is 
%strange since we say that inferring all 38 takes 4 hours 
%but here 1 network takes 5 hours.

\begin{figure*}[t]
	\centering
	\subfigure[Fukushima (2011-03-18)]{\includegraphics[width=0.32\textwidth]{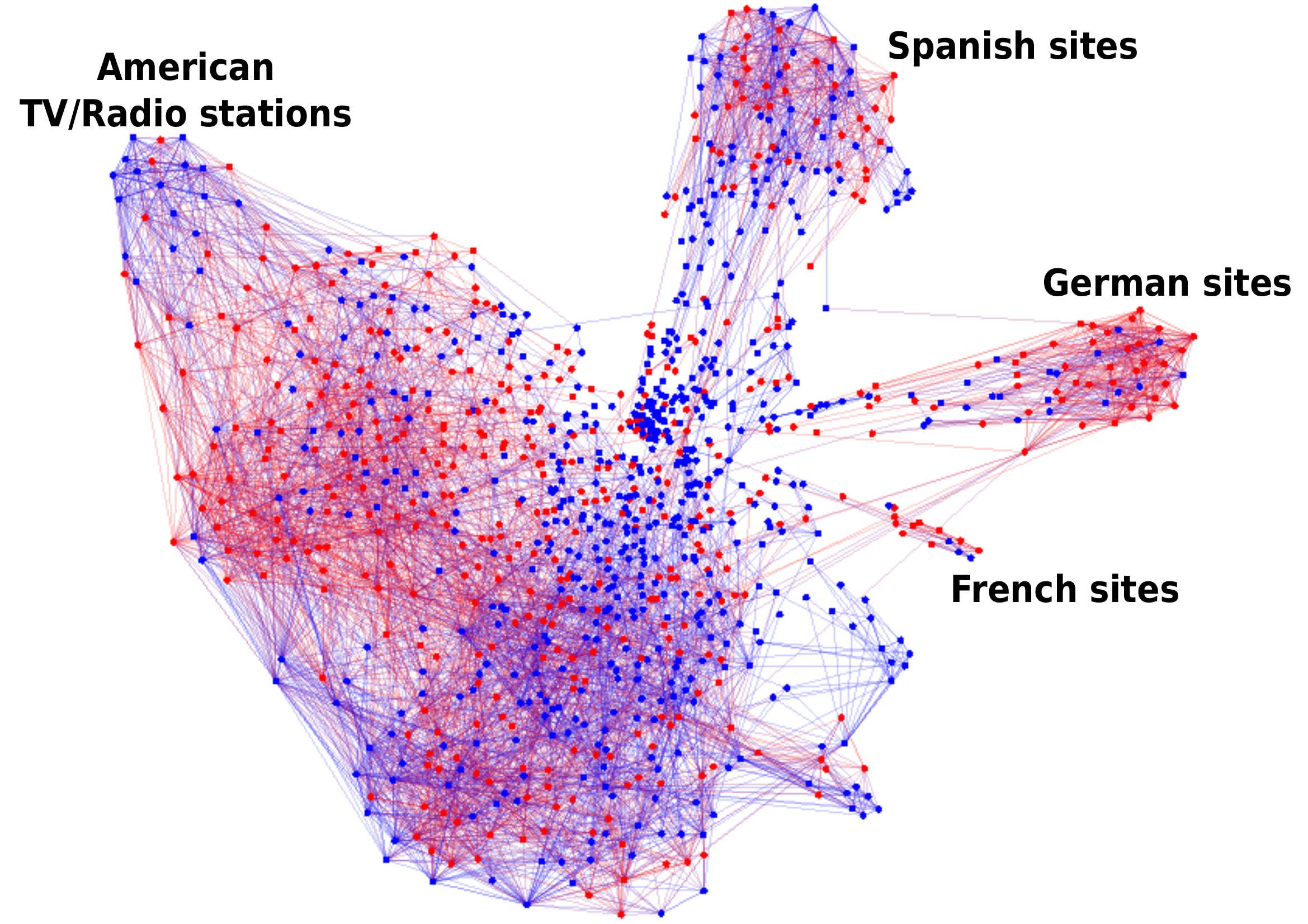}
	\label{fig:network-fukushima-2011-03-18}}
	\subfigure[Fukushima (2011-06-25)]{\includegraphics[width=0.32\textwidth]{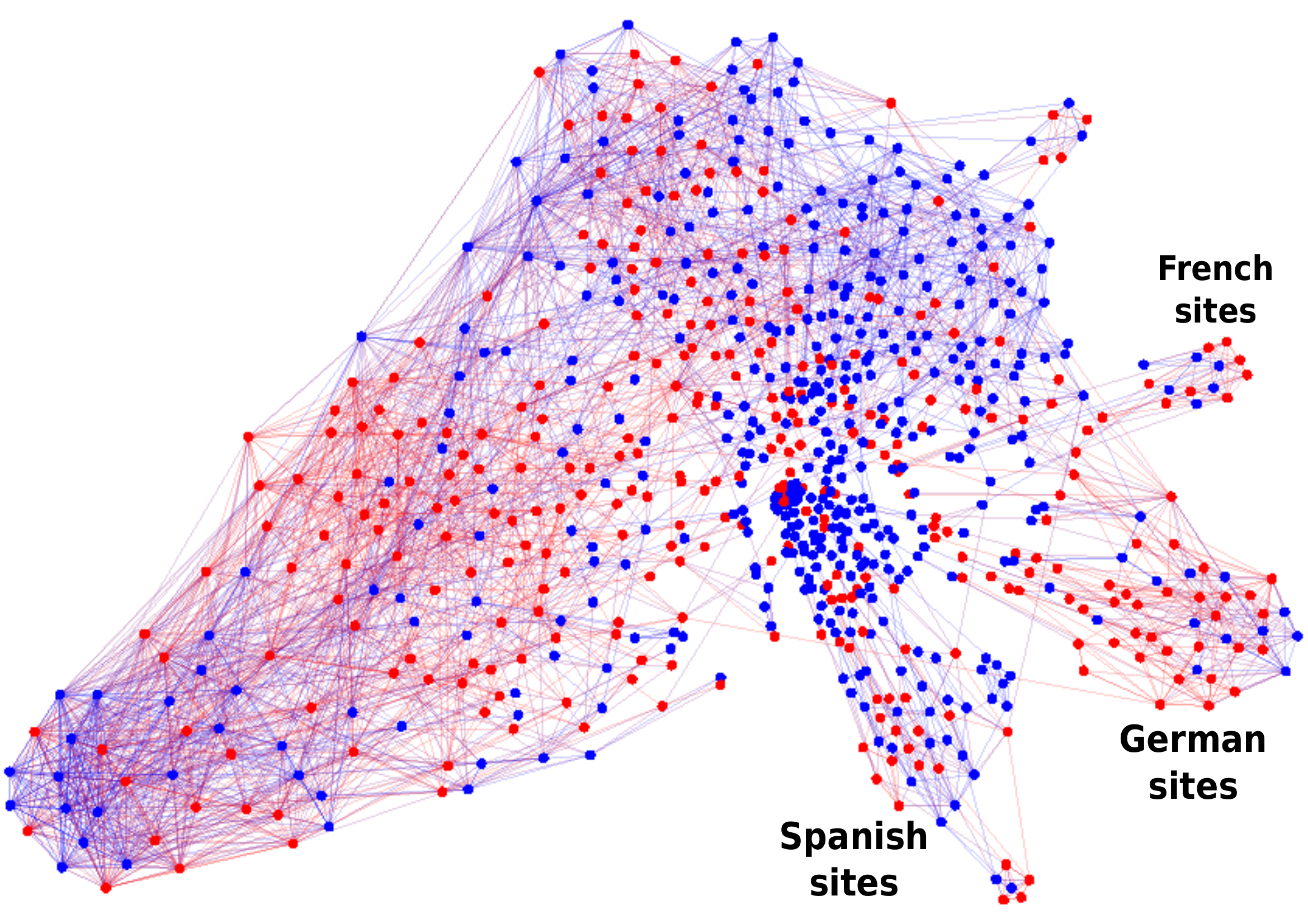}
	\label{fig:network-fukushima-2011-06-25}}
	\subfigure[Fukushima (2011-10-13)]{\includegraphics[width=0.32\textwidth]{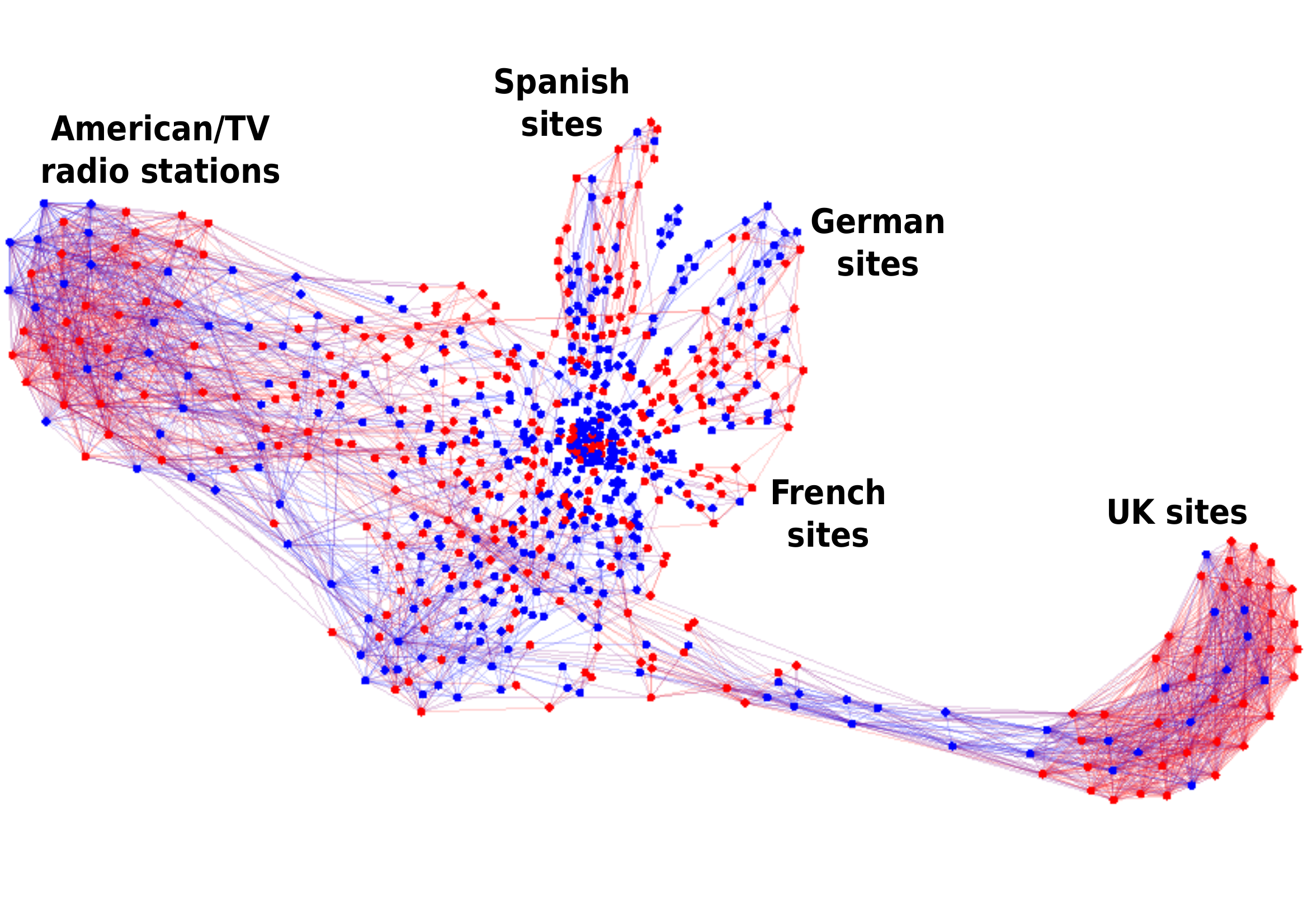}
	\label{fig:network-fukushima-2011-10-13}} \\
	\subfigure[UK royal wedding (2011-04-02)]{\includegraphics[width=0.32\textwidth]{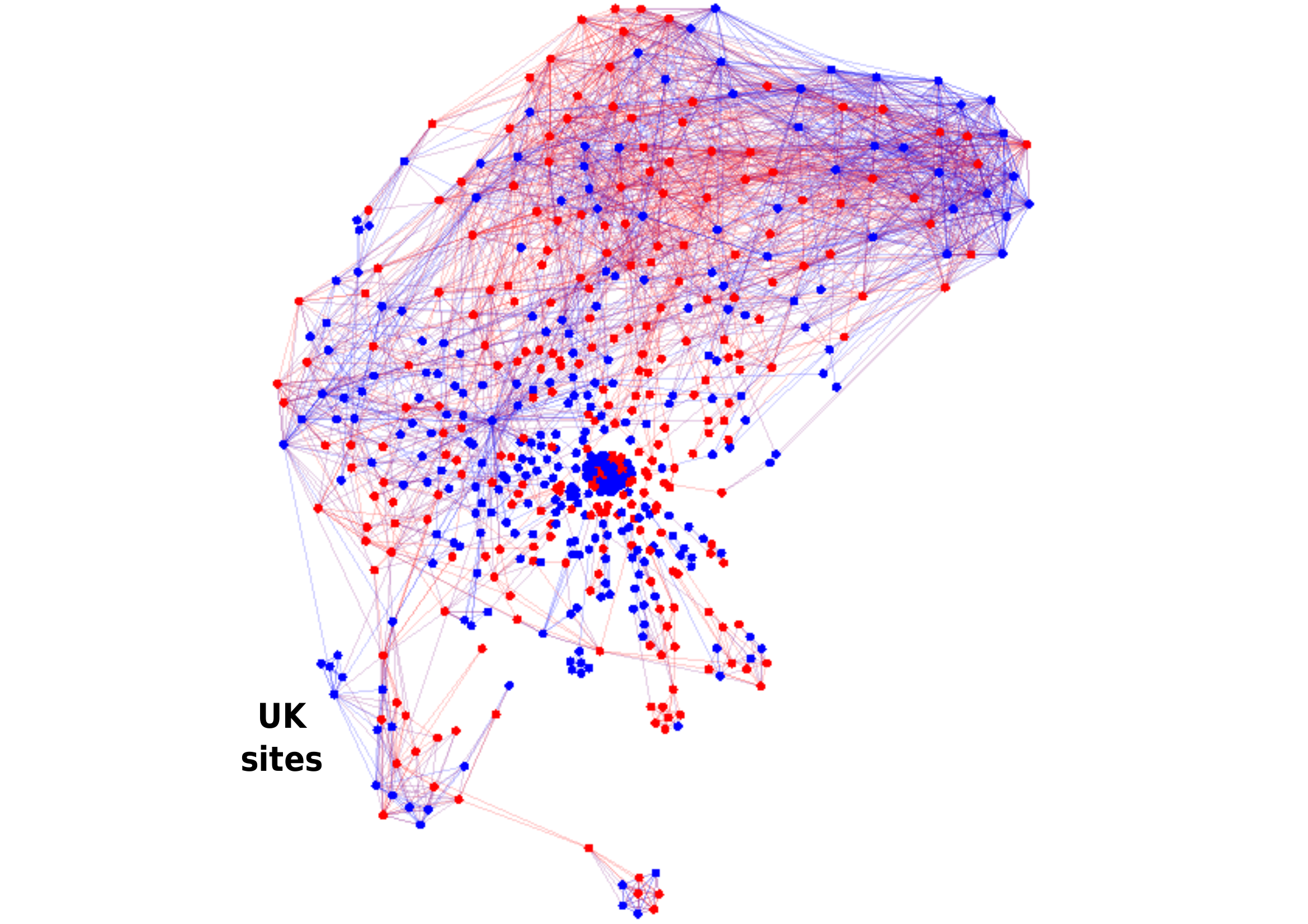}
	\label{fig:network-middleton-2011-04-02}}
	\subfigure[UK royal wedding (2011-05-02)]{\includegraphics[width=0.32\textwidth]{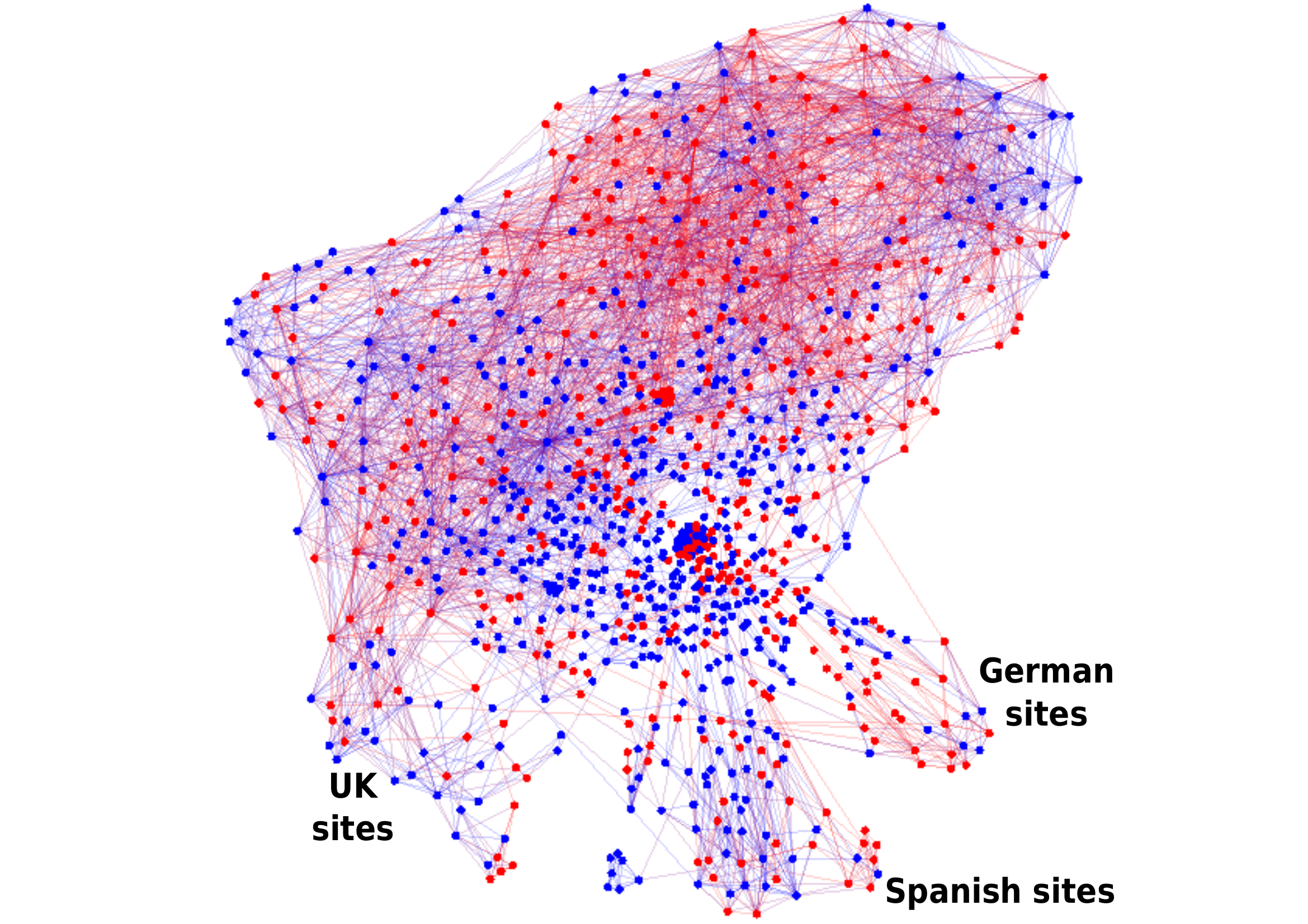}
	\label{fig:network-middleton-2011-05-02}}
	\subfigure[UK royal wedding (2011-11-15)]{\includegraphics[width=0.32\textwidth]{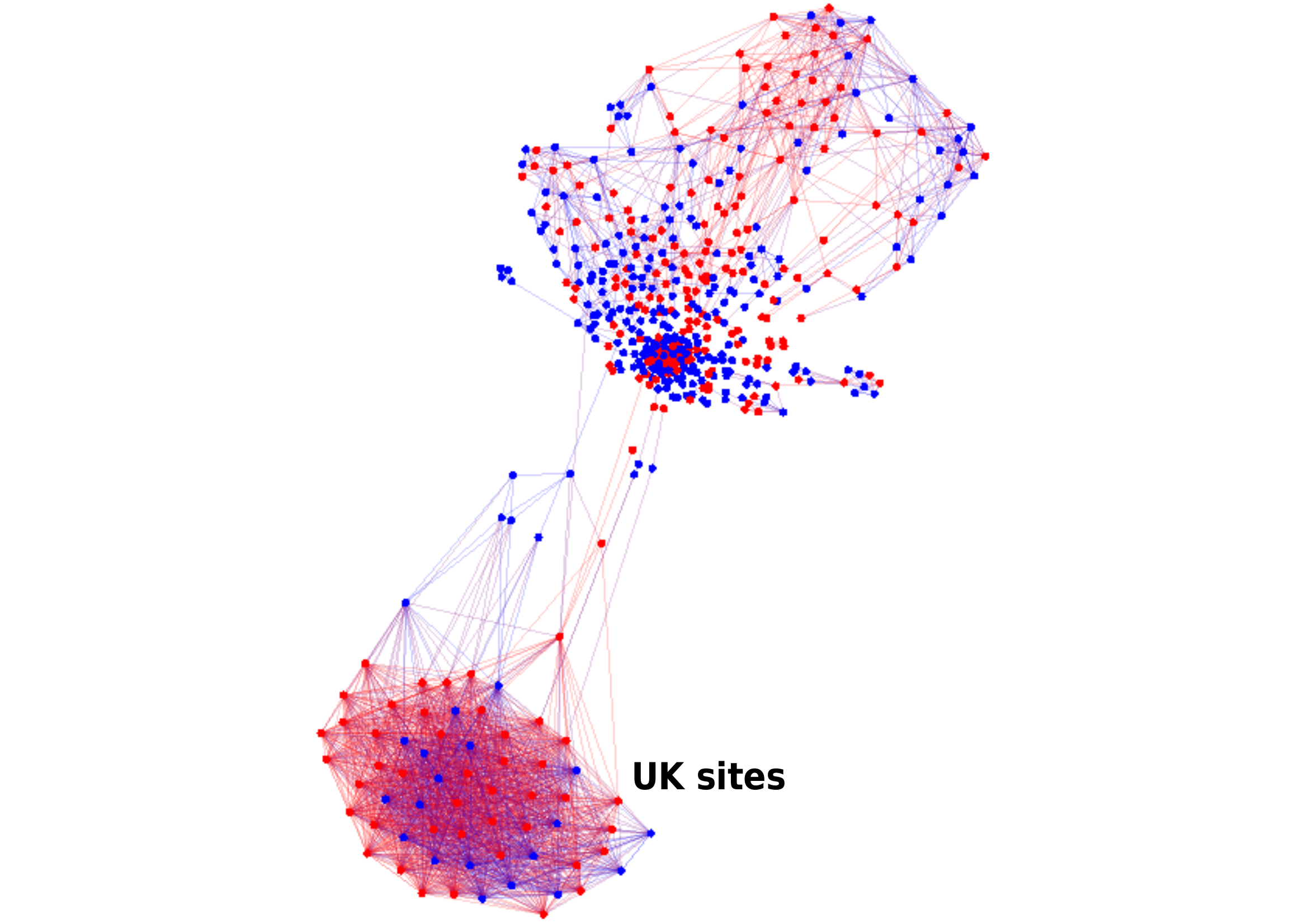}
	\label{fig:network-middleton-2011-11-15}} \\
	\subfigure[Syria'{}s uprise (2011-04-05)]{\includegraphics[width=0.32\textwidth]{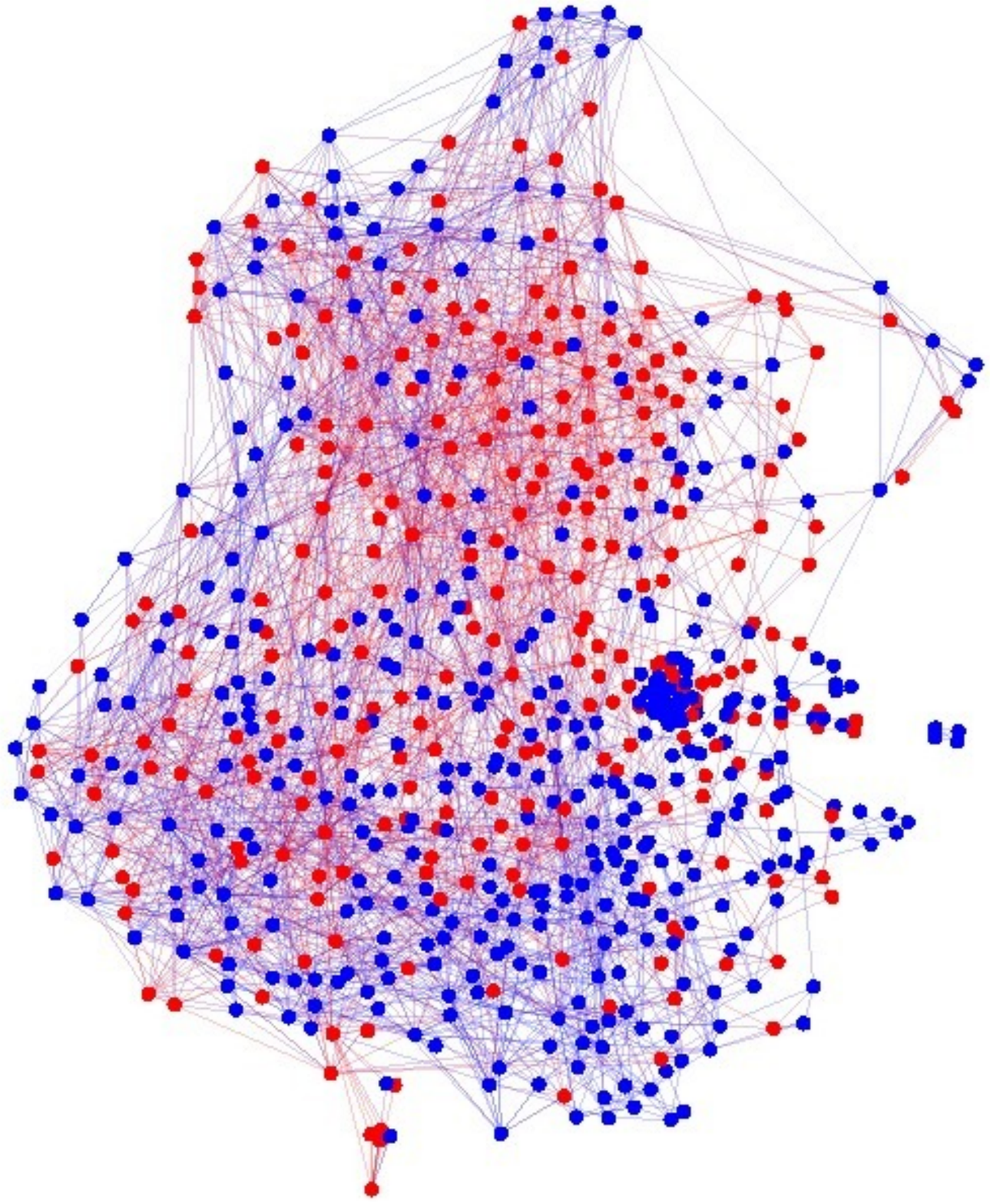}
	\label{fig:network-syria-2011-04-05}}
	\subfigure[Syria'{}s uprise (2011-06-02)]{\includegraphics[width=0.32\textwidth]{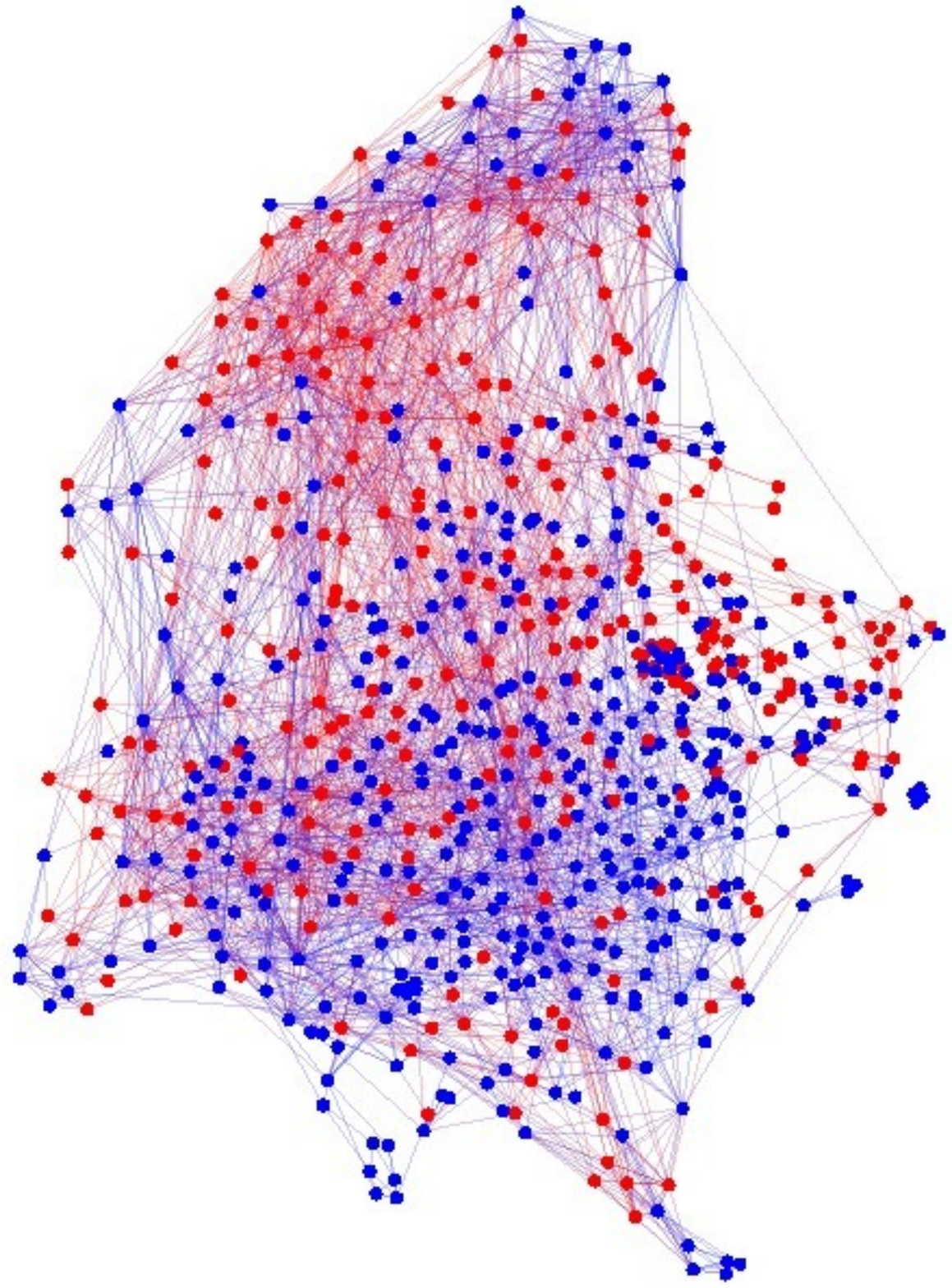}
	\label{fig:network-syria-2011-06-02}}
	\subfigure[Syria'{}s uprise (2011-12-02)]{\includegraphics[width=0.32\textwidth]{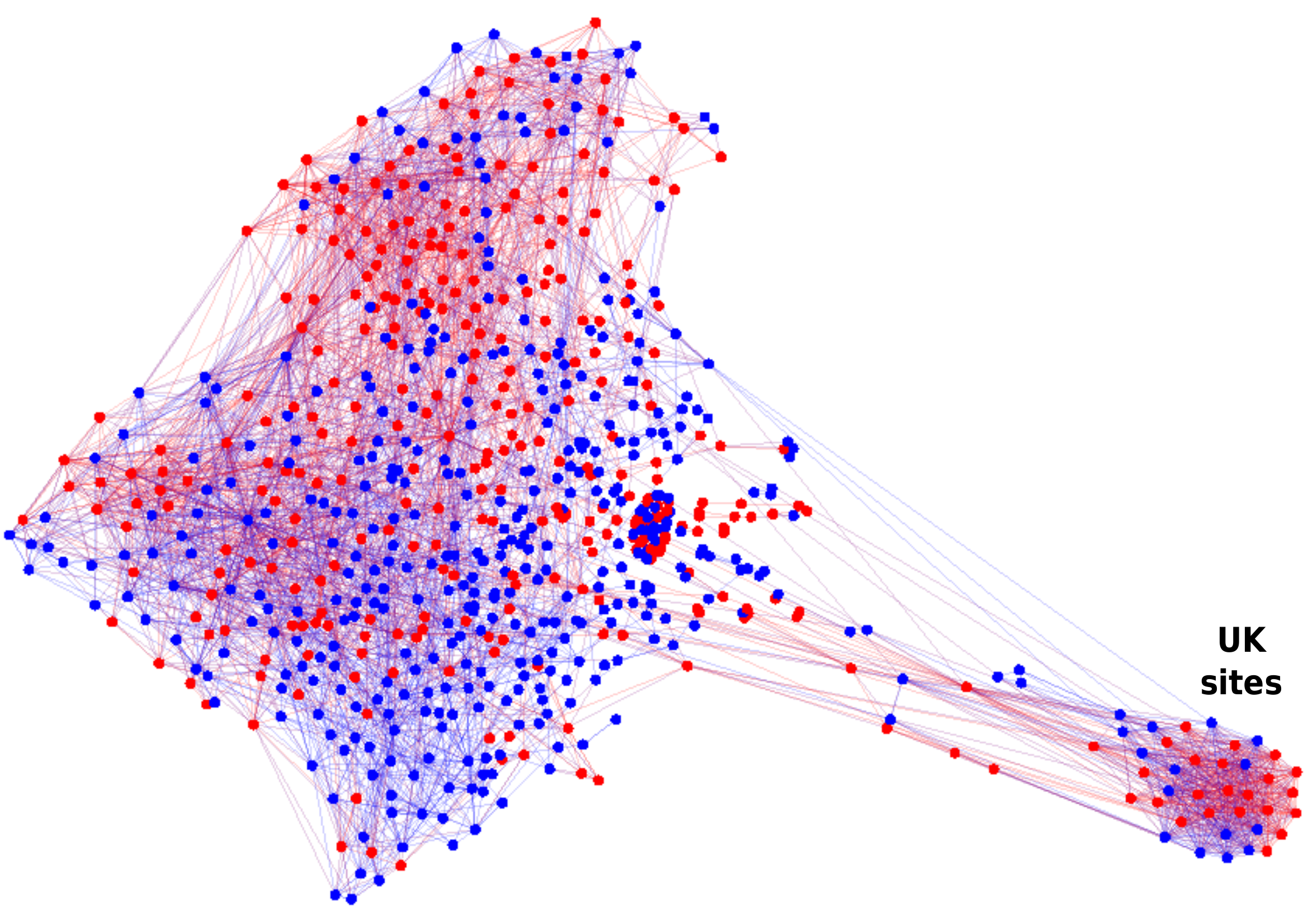}
	\label{fig:network-syria-2011-12-02}}
	\vspace{-3mm}
 	\caption{Time-varying diffusion networks for three different major events of 2011. Red nodes are mainstream media, and blue nodes are blogs. Additional visualizations for other topics and events are available at the supporting website~\cite{website13}.} \label{fig:networks}
 	\vspace{-3mm}
\end{figure*}

\xhdr{Visualizing the information pathways}
Figure~\ref{fig:networks} plots diffusion networks for three different 2011 world events: Fukushima nuclear disaster, UK royal wedding, and civil uprise in Syria. Each network is shown at three different time points. Red nodes represent mainstream media sites, and blue nodes represent blogs~\cite{leskovec2009kdd}. 

Based on the figure, we draw several interesting observations. Most often, information propagates through a core-periphery network structure. Such structure emerges by few central media sites and blogs driving the adoption of memes across the Web~\cite{manuel10netinf}.
However, the network structure often changes dramatically over time, and we find clusters that emerge and vanish in short periods of time. For example, the information networks for Syria'{}s uprise illustrated in Figures~\ref{fig:networks}(g-h), do not have any clear clustering structure. However, on December 2, 2011 (Figure~\ref{fig:network-syria-2011-12-02}) a cluster suddenly emerges in the network. Further investigation reveals that the cluster is composed of UK news sites and blogs that discuss recently implemented EU sanctions against Syria.
Generally, it is common to observe sudden formation of clusters of sites from specific geographical areas. This is specially noticeable in the information network for Fukushima{}'s disaster, in Figures~\ref{fig:networks}(a-c). Such clusters often form due to language boundaries, since such boundaries prevent memes to flow across countries or continents. Moreover, we often observe that such clusters are caused by a common external event~\cite{seth2012kdd}, like in the case of UK discussion on EU sanctions against Syria. Inferred dynamic networks can thus be used to investigate the flow of information as well as to detect external events that cause sudden perturbations to the diffusion network structure.

\begin{table}[!!t]
    \small
    \begin{center}
    \begin{tabular*}{0.48\textwidth}{@{\extracolsep{\fill}} l c c}
	\toprule
  \textbf{Topic or news event $(\bf{Q})$} & \textbf{\# sites} & \textbf{\# memes} \vspace{1mm} \\
	\midrule
	Amy Winehouse & 1,207 & 109,650 \\
	\midrule
	Fukushima & 1,666 & 383,745 \\
	\midrule
	Gaddafi & 1,358  & 440,646 \\
	\midrule
	Kate Middleton & 1,427 & 191,777 \\
	\midrule
	NBA & 2,087 & 1,543,630\\
	\midrule
	Occupy & 1,875 & 655,183 \\
	\midrule
	Strauss-Kahn & 1,263 & 204,238 \\
	\midrule
	Syria & 1,565 & 615,176 \\
	\bottomrule
    \end{tabular*}
    \end{center}
	\vspace{-4mm}
  \caption{Topic and news event statistics.}
  \label{tab:cascades}
  \vspace{-5mm}
\end{table}

\begin{figure*}[t]
	\centering
	\subfigure[Amy Winehouse]{\includegraphics[width=0.23\textwidth]{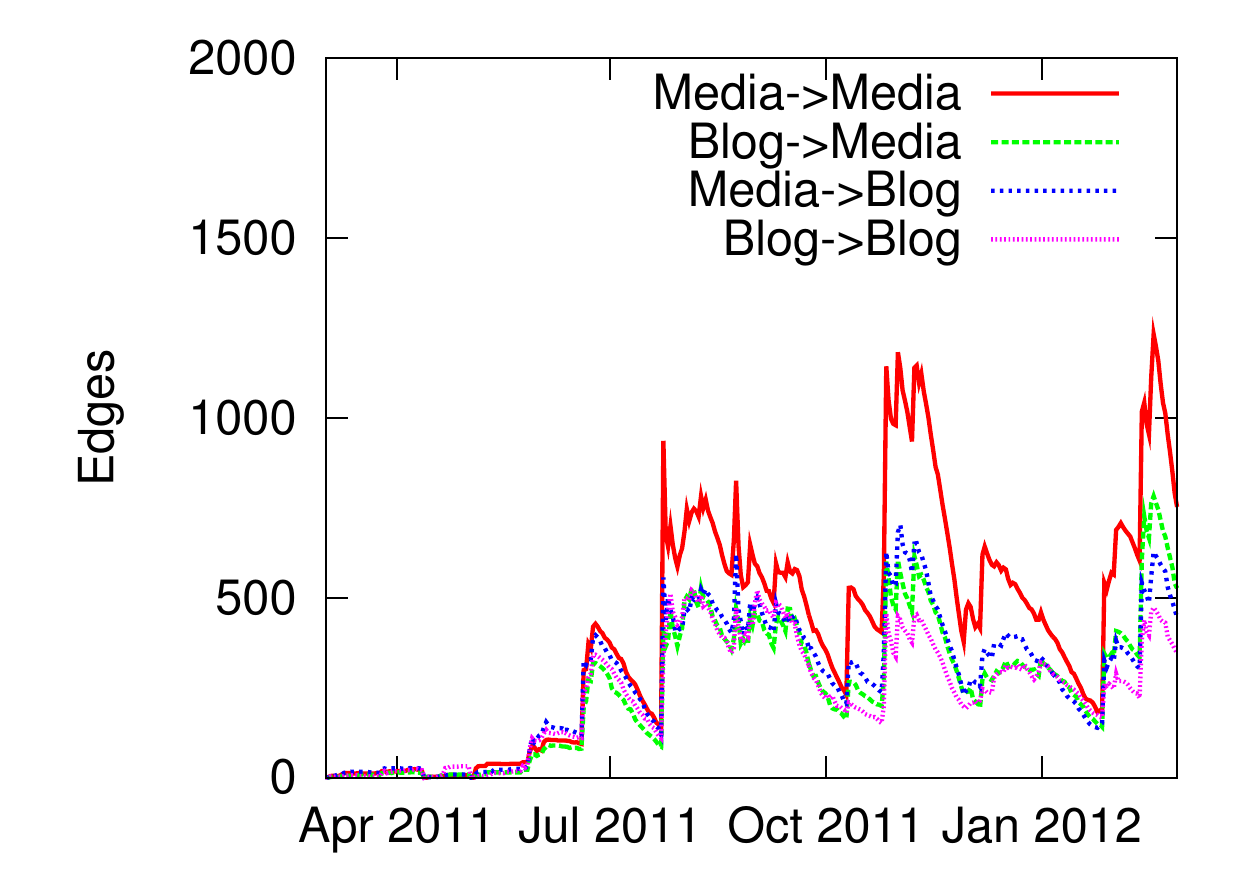}
	\label{fig:type-edges-amy-winehouse}}
	 \subfigure[Fukushima]{\includegraphics[width=0.23\textwidth]{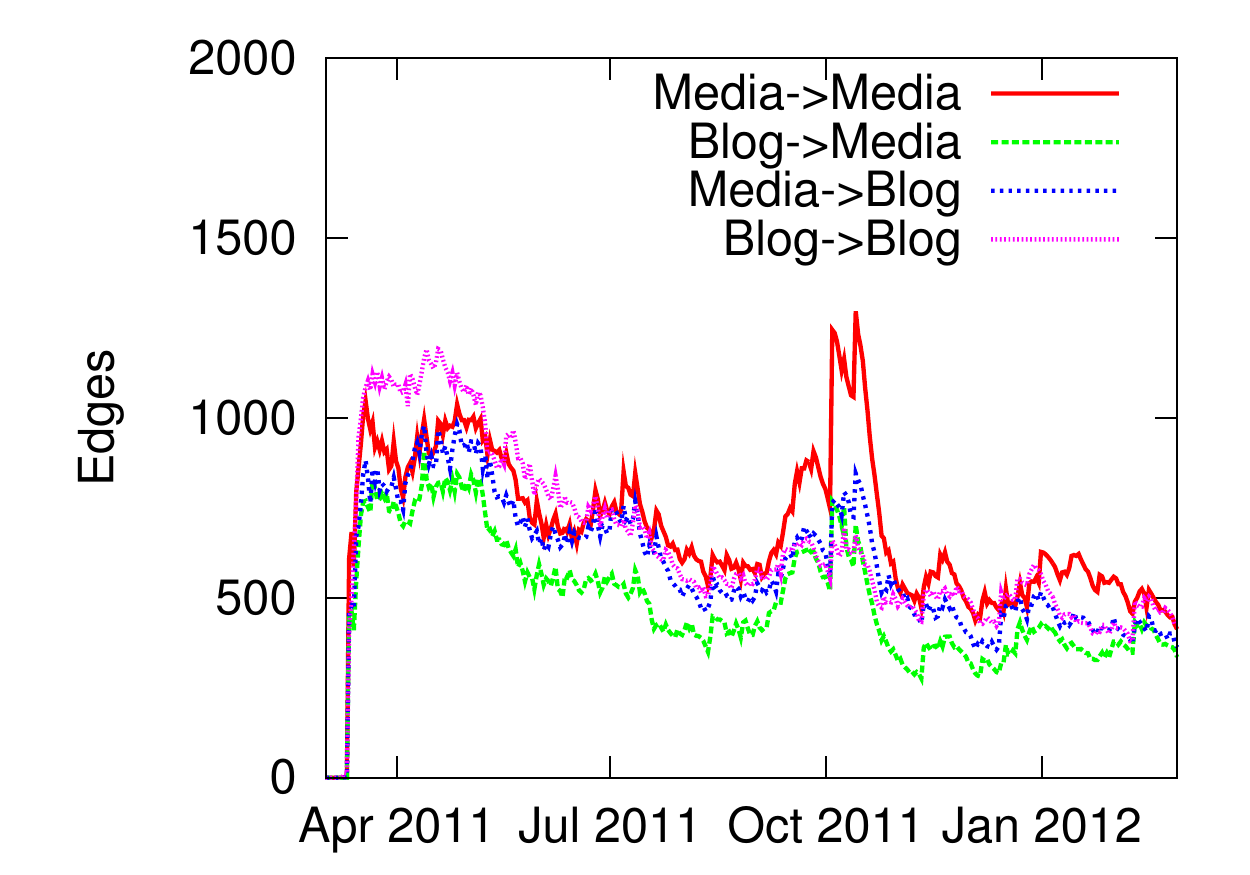}
	\label{fig:type-edges-fukushima}}
	\subfigure[Gaddafi]{\includegraphics[width=0.23\textwidth]{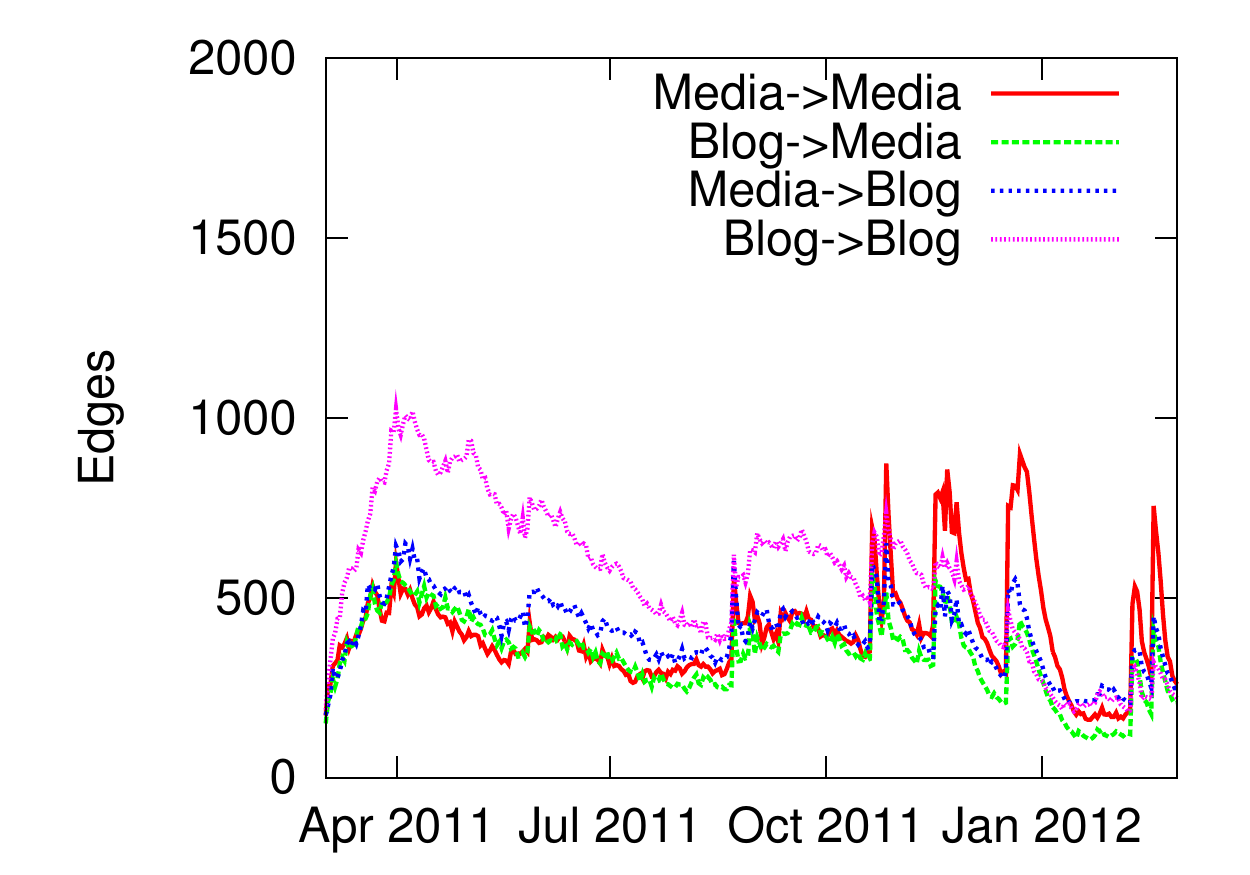}
	\label{fig:type-edges-gaddafi}}
	\subfigure[UK royal wedding]{\includegraphics[width=0.23\textwidth]{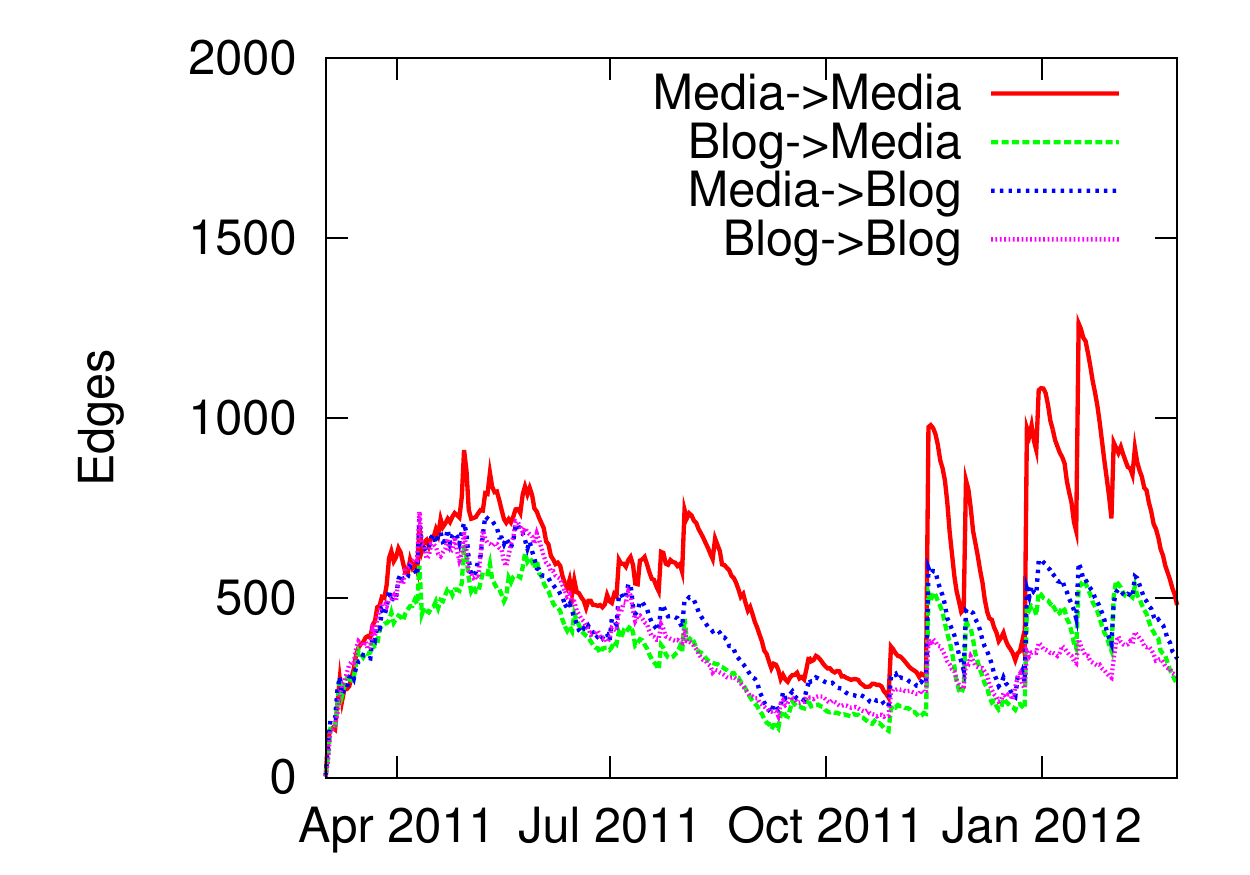}
	\label{fig:type-edges-middleton}} \\
	\subfigure[NBA]{\includegraphics[width=0.23\textwidth]{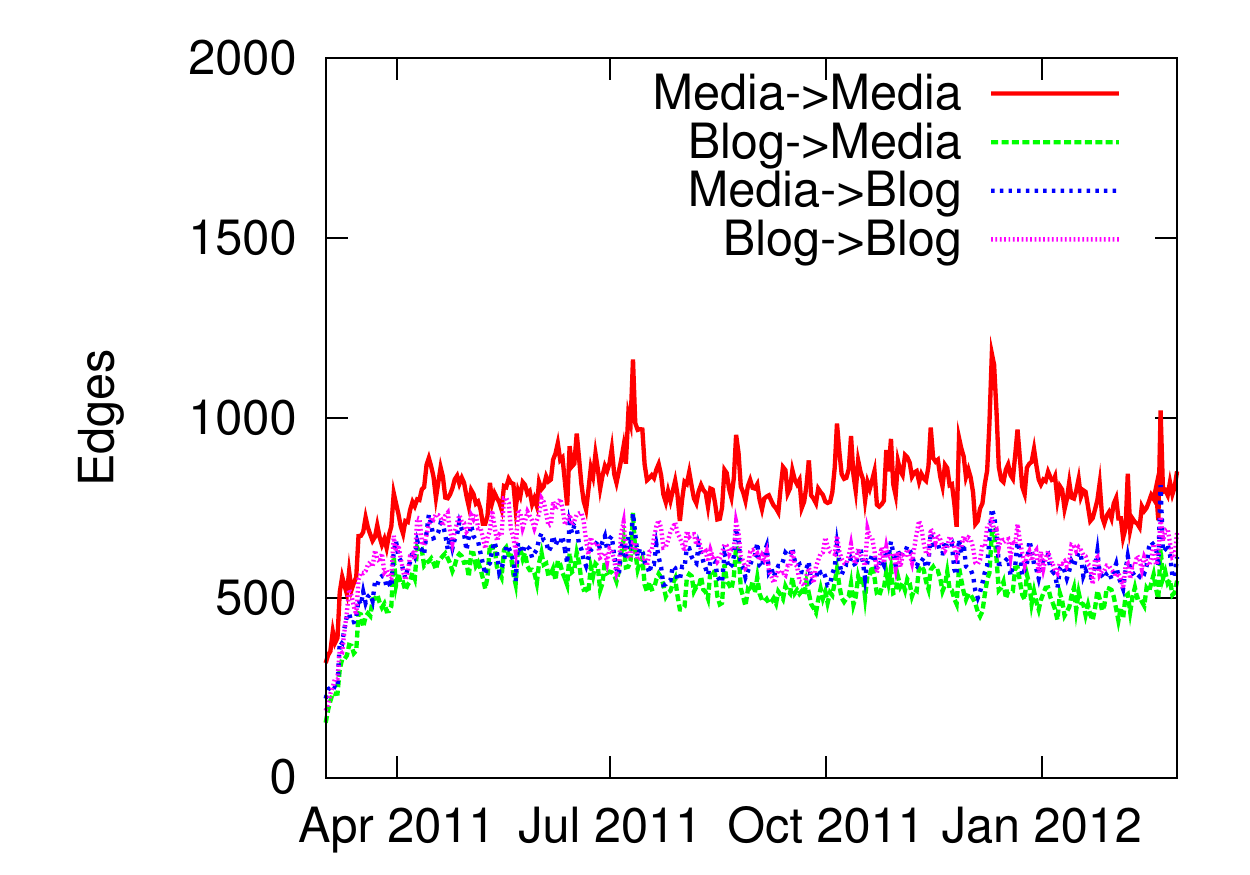}
	\label{fig:type-edges-nba}}
	\subfigure[Occupy]{\includegraphics[width=0.23\textwidth]{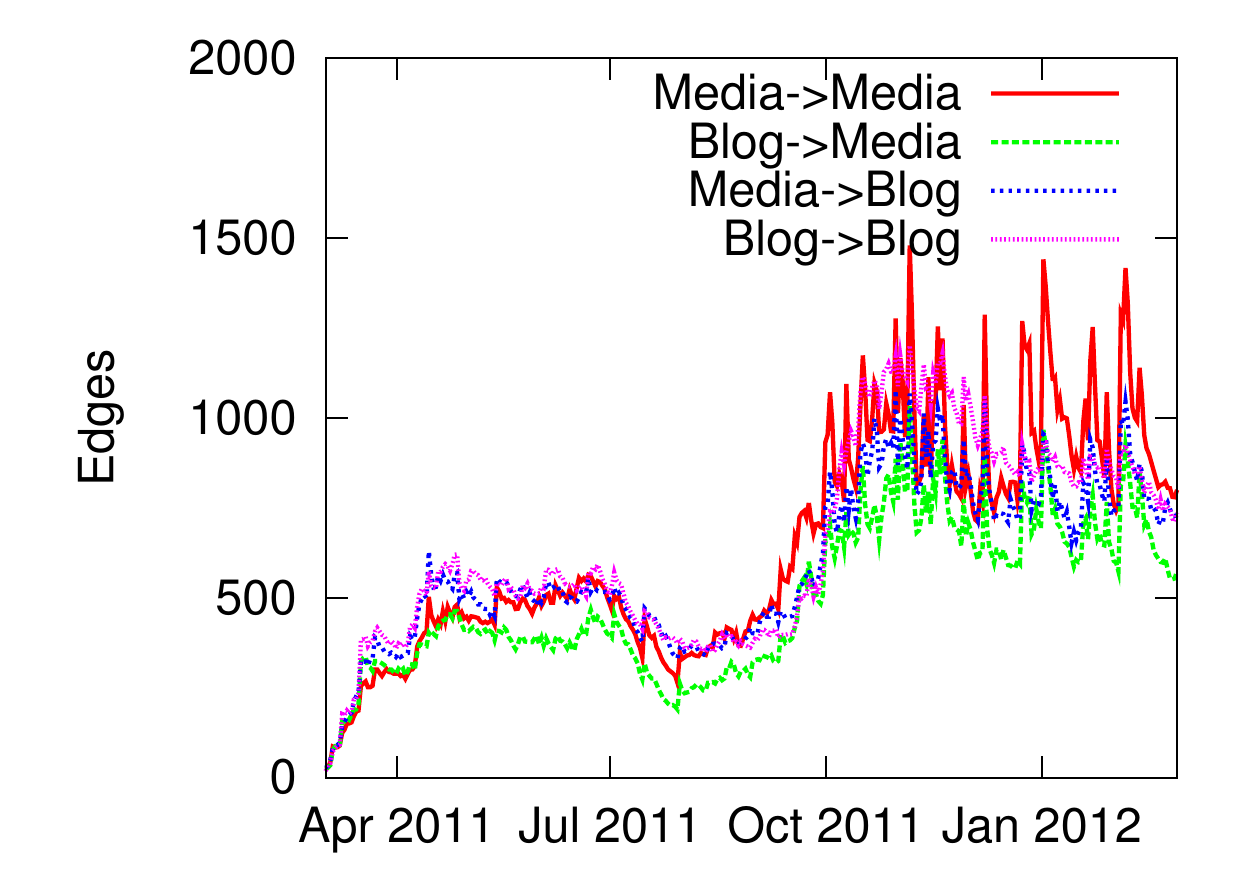}
	\label{fig:type-edges-occupy}}
	 \subfigure[Strauss-Kahn]{\includegraphics[width=0.23\textwidth]{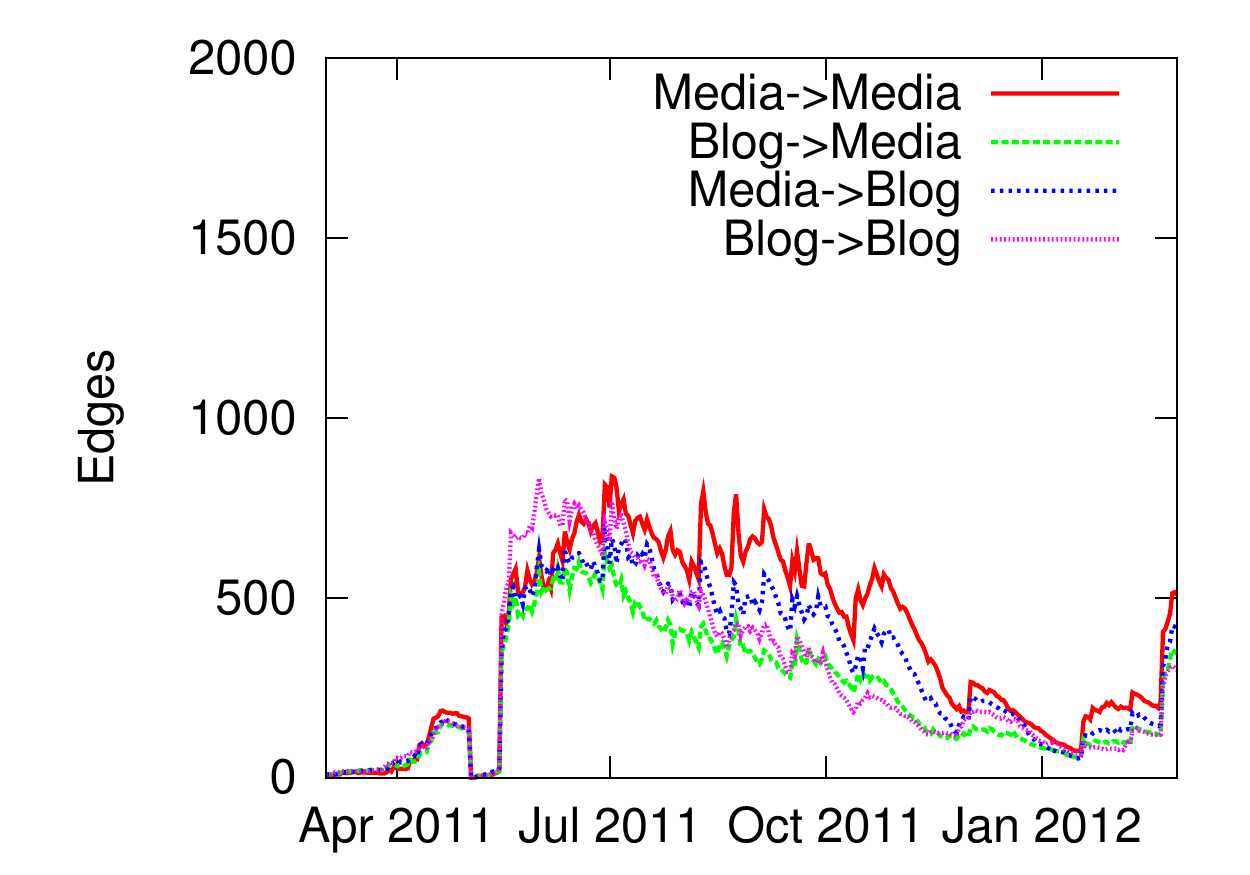}
	\label{fig:type-edges-strauss-kahn}}
	\subfigure[Syria]{\includegraphics[width=0.23\textwidth]{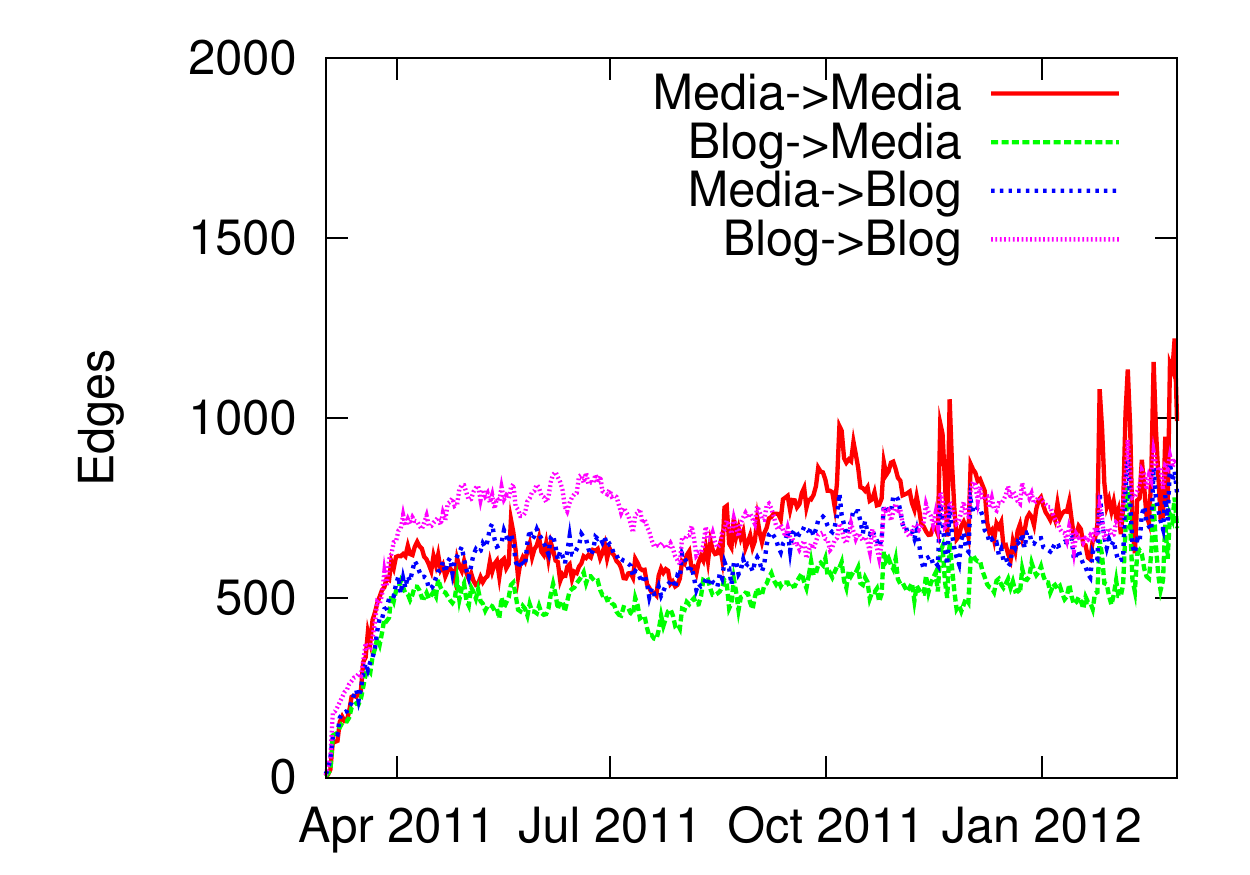}
	\label{fig:type-edges-syria}}
	%\vspace{-3mm}
 	\caption{The number of links that point between different types of sites over time for eight different inferred diffusion networks. We split the sites into mainstream media and blogs and count the links among these two node types.} \label{fig:type-edges}
 	%\vspace{-3mm}
\end{figure*}

\xhdr{Evolution of edge transmission rates} Next, we aim to study the evolution of links among different types of sites. We label the nodes in our network as mainstream media and blog, and compute the number of links between different types of sites over time. Figure~\ref{fig:type-edges} gives the results for several inferred diffusion networks for different topics and world events.  We note several interesting patterns.

The connectivity changes tend to reflect the amount of attention that a news event or a topic triggers over time. Unexpected news events, like the sex scandal of the director of the International Monetary Fund Strauss-Kahn on May 14, 2011
in Fig.~\ref{fig:type-edges-strauss-kahn} or the death of British singer Amy Winehouse on July 23, 2011 in Fig.~\ref{fig:type-edges-amy-winehouse}, result in a dramatic increase in the number of edges over a short period of time. More general topics, like the NBA in Fig.~\ref{fig:type-edges-nba}, result in a network with more stable connectivity over time.
Certain types of news are sometimes spreading earlier among blogs than mainstream media. This is especially the case for population wide events like the Fukushima nuclear disaster, civil war in Libya and civil uprise in Syria (Fig.~\ref{fig:type-edges}(b, c, h). However, it happens more frequently that the largest amount of links are mainstream media-to-mainstream media and the fewest links point from blogs to mainstream media. These results are intuitive and consistent with previous work~\cite{manuel10netinf,leskovec2009kdd} that observed most often information flows from mainstream media to blogs (and rarely the other way around). However, as we see here for population level events and social movements (like, in case of the civil unrest in the Middle East) social media plays crucial role in information dissemination and organization of civil movements.

\xhdr{Evolution of node centrality} Having studied the dynamics of edges in the network we now move towards investigating the network centrality of blogs and mainstream media sites over time for different topics and world events. 
%and compute the percentage of mainstream media and blogs among the most central sites. 
To measure network centrality of node $S$ in the network at time $t$, we first compute shortest path length from $S$ to any other node $R$ in the network. Then centrality of node $S$ is defined as $\sum_R 1/d(S, R)$, where $d(S, R)$ is the shortest path length from $S$ to $R$ (if $R$ is not reachable from $S$ then $d(S, R)=\infty$). 
For networks with core-periphery structure, nodes with high centrality are typically located in the ``central'' core of the network.

%run breadth-first search over the inferred network $\hat{G}(t)$ and count the number of reachable
%sites from $S$, 

\begin{figure*}[t]
	\centering
	\subfigure[Amy Winehouse]{\includegraphics[width=0.23\textwidth]{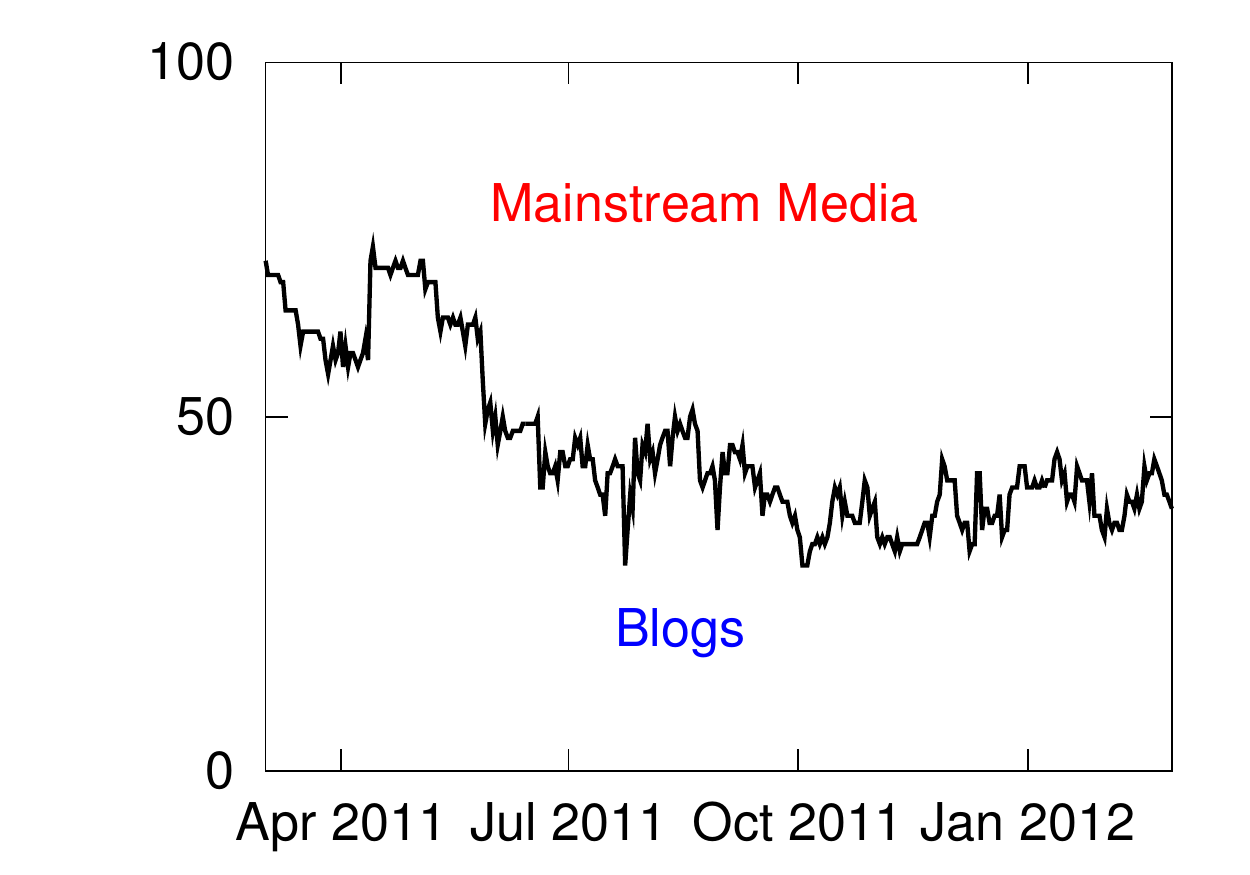}
	\label{fig:reachability-amy-winehouse}}
	\subfigure[Fukushima]{\includegraphics[width=0.23\textwidth]{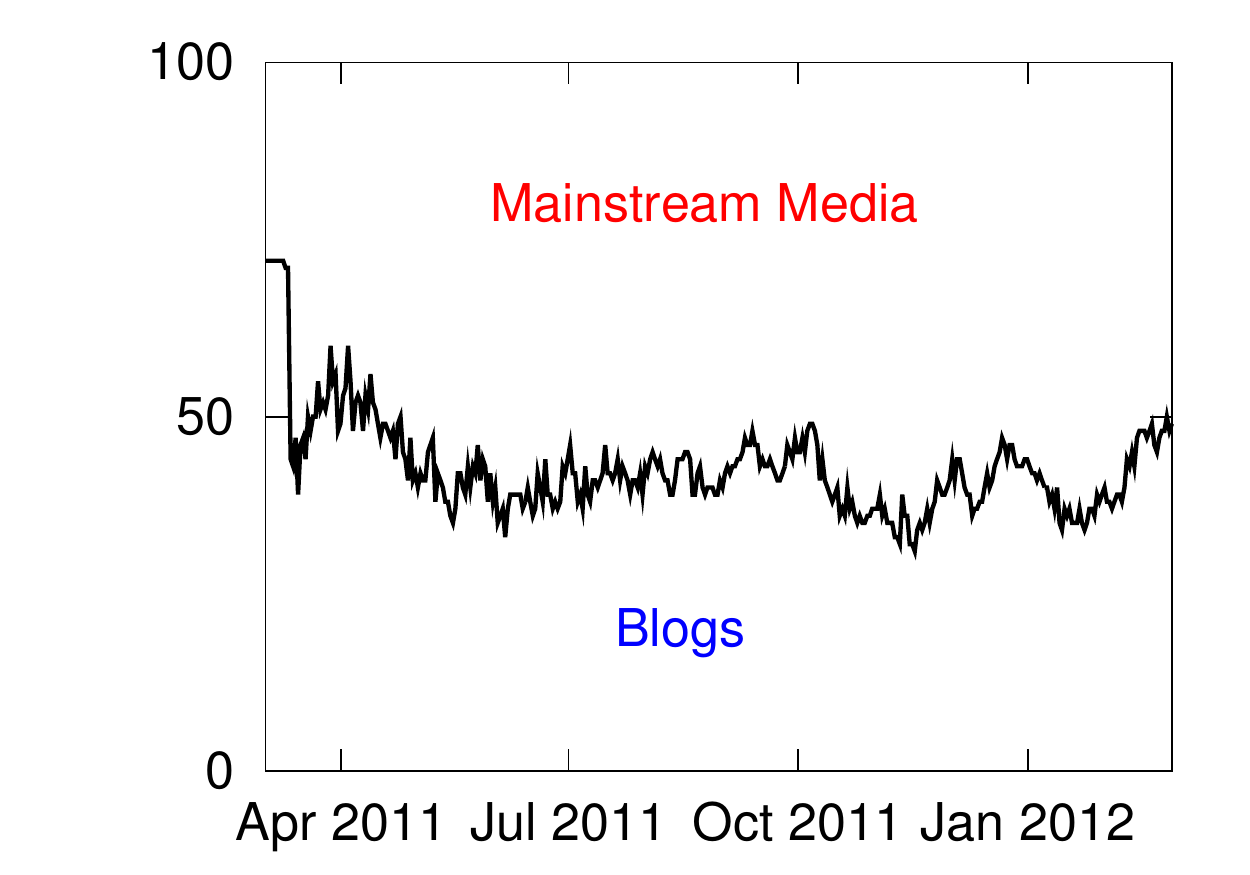}
	\label{fig:reachability-fukushima}}
	\subfigure[Gaddafi]{\includegraphics[width=0.23\textwidth]{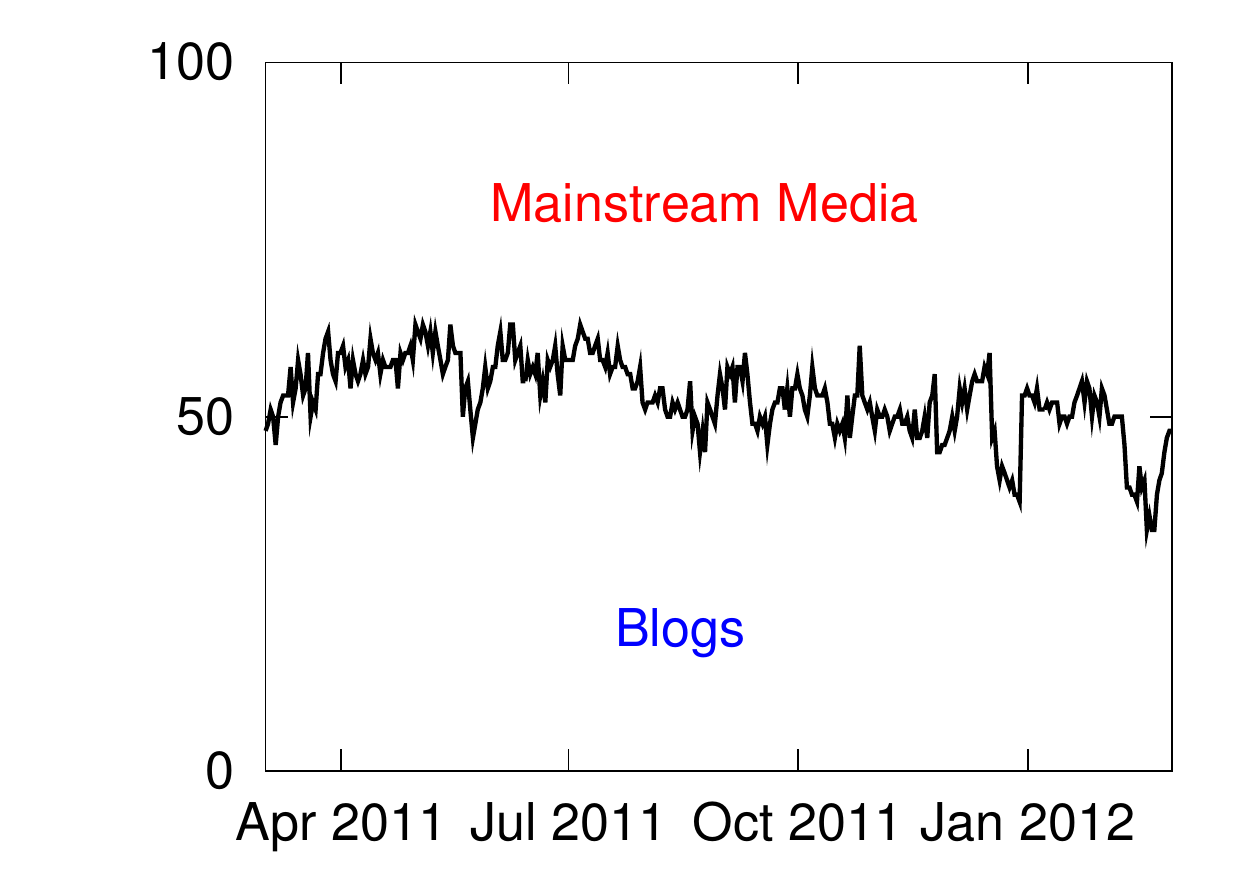} 
	\label{fig:reachability-gaddafi}}
	\subfigure[UK royal wedding]{\includegraphics[width=0.23\textwidth]{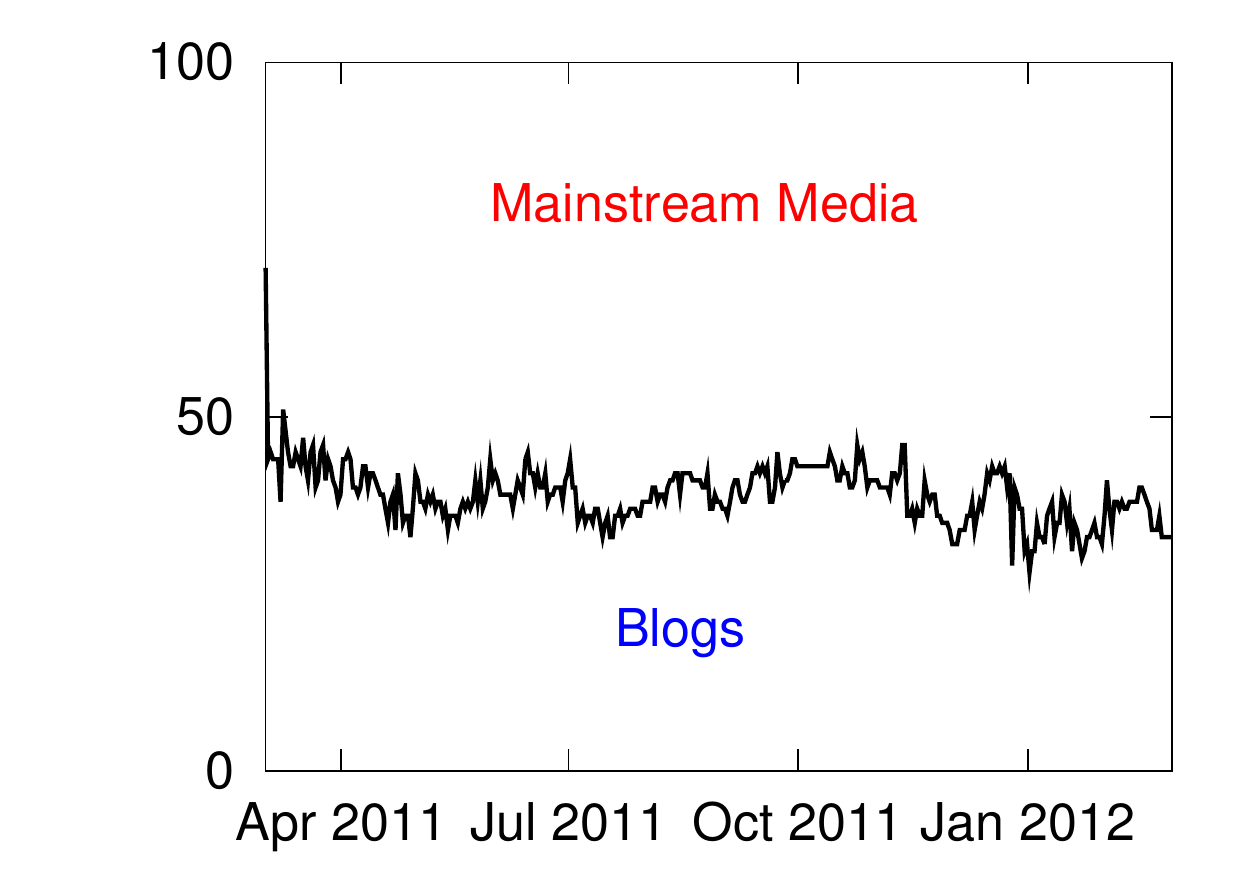} 
	\label{fig:reachability-middleton}}
	 \\
	\subfigure[NBA]{\includegraphics[width=0.23\textwidth]{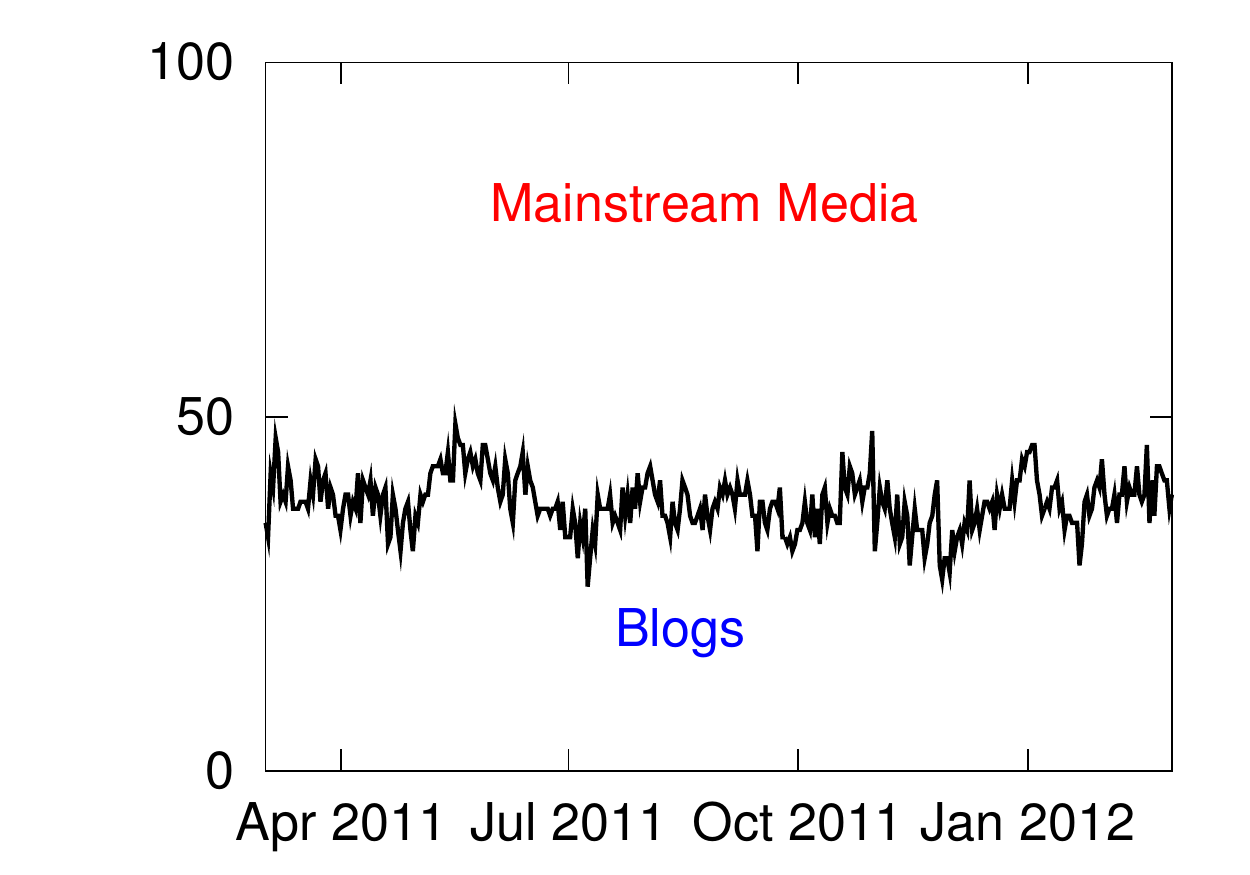}
	\label{fig:reachability-nba}}
	\subfigure[Occupy]{\includegraphics[width=0.23\textwidth]{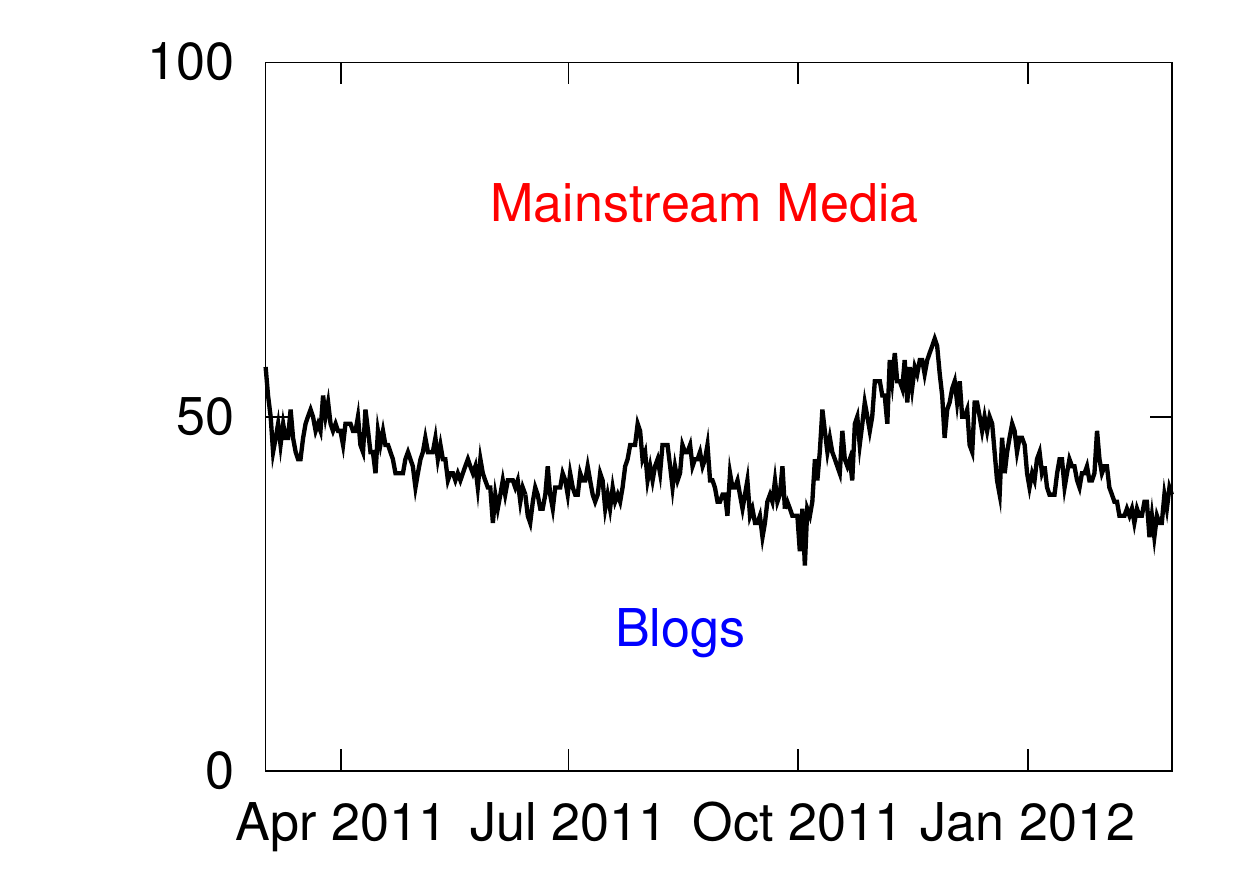}
	\label{fig:reachability-occupy}}
	\subfigure[Strauss-Kahn]{\includegraphics[width=0.23\textwidth]{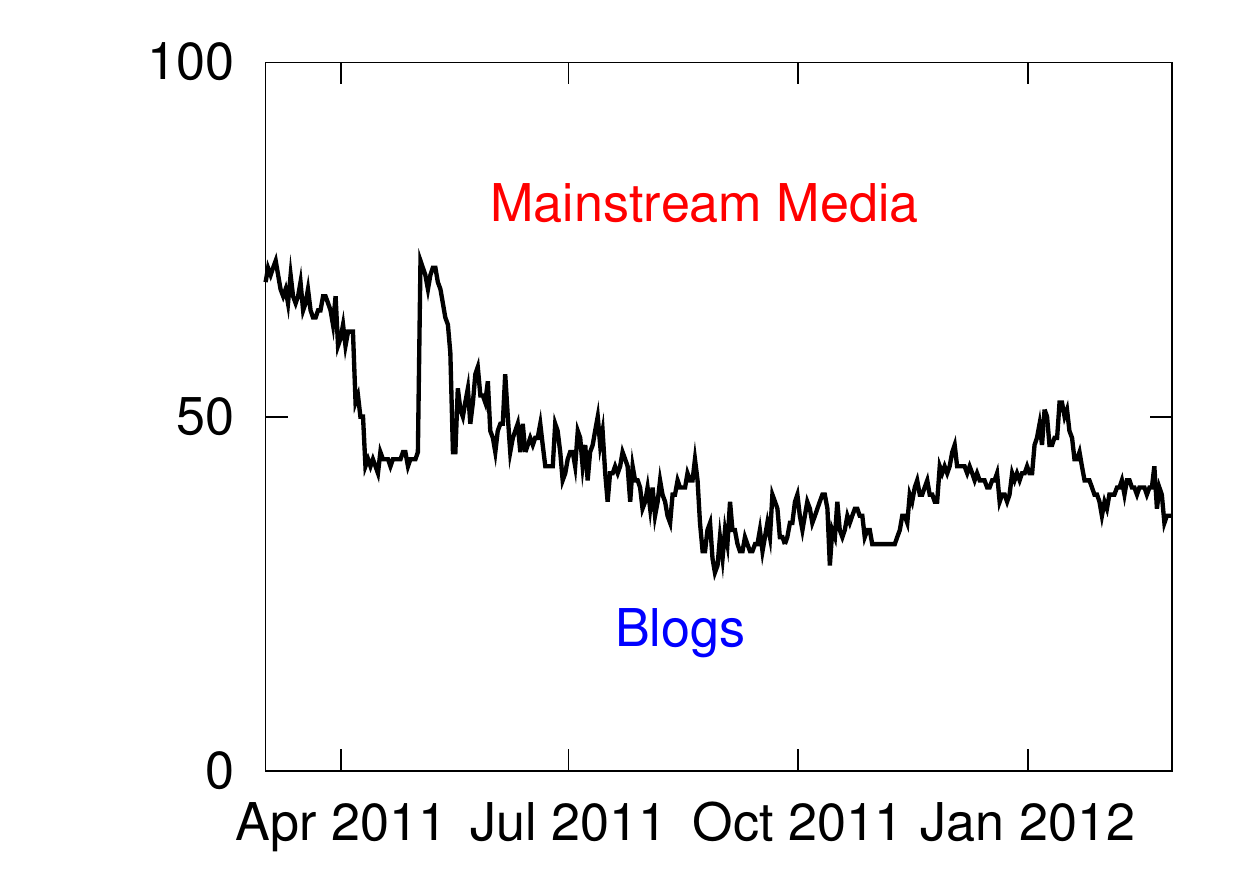}
	\label{fig:reachability-strauss-kahn}}
	\subfigure[Syria]{\includegraphics[width=0.23\textwidth]{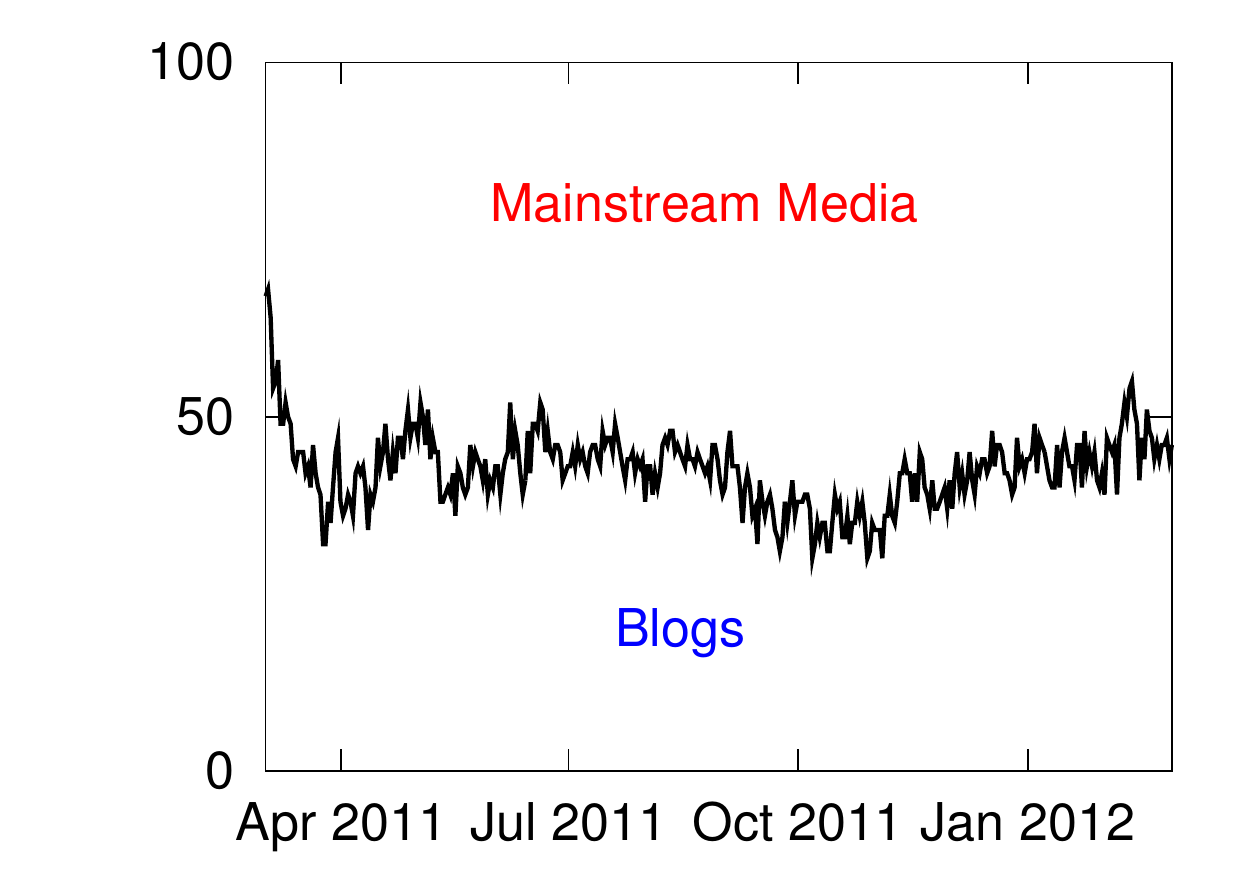}
	\label{fig:reachability-syria}}
	\vspace{-3mm}
 	\caption{Percentage of blogs and mainstream media in top-100 most \emph{central} sites for eight different inferred diffusion networks. Mainstream media
	are represented in red, and blogs in blue.} \label{fig:reachability}
	\vspace{-3mm}
 	\end{figure*}

Figure~\ref{fig:reachability} plots the percentage of blogs among the top 100 most central sites over time for eight different topics/events of 2011. Perhaps surprisingly, we observe there is a about the same number of mainstream media and blogs in the top-100 most central nodes for most networks -- the number of blogs in the top-100 does not typically
decreases below 30\% or increases over 70\%.
For some topics, mainstream media are always more \emph{central} (\eg, baseball and NBA in Figures~\ref{fig:reachability}(a, b)). In contrast, for other topics, blogs dominate mainstream media over a significant amounts of time (\eg, Gaddafi in Fig.~\ref{fig:reachability-gaddafi}).
Centrality of mainstream media and blogs can be relatively constant (Fig.~\ref{fig:reachability}(a,b)) or more time-varying (Fig.~\ref{fig:reachability}(c,h)). We find that a significant rise in the number of central blogs is often temporally correlated with an increasing social unrest (\eg, the Occupy Wall Street movement in Sept-Nov 2011 in Fig.~\ref{fig:reachability-occupy}).

\xhdr{Accuracy on real data} So far, we have used \emph{memes} to trace the flow of information over the Web and have made several qualitative observations about the structure and dynamics of information pathways in online media.  %However, they do not allow us to evaluate quantitatively our method since we cannot build a ground truth from the memes.
We now proceed and attempt to also quantitatively evaluate \infopath on real data. In case of real data the ground-truth information diffusion network is impossible to obtain. However, we can use the temporal dynamics of hyperlinks created between news sites as a proxy for real information flow. Thus, by observing the times when sites create hyperlinks, our goal is to infer the `targets' of the links (i.e., infer the hyperlink network from the hyperlinks times).

%to evaluate the accuracy of our algorithm in real data, since they allow us to build a time-varying groundtruth network $G^*(t)$, although they are
%much less frequent than \emph{memes}, specially among mainstream media. 
%
We proceed as follows. 
First, we discretize the time in days, we generate one network $G^*(t)$ per day $t$, in which we add an edge $(u,v)$ if a document on a site $u$ linked to a document on a site $v$ within the last day.
Then, we build a set of hyperlink cascades. A hyperlink cascade $c_h$ starts when a site publishes a piece of information and then other sites use hyper-links to refer to it. 
% These other sites link to still others and so on. 
Since all our documents/posts are time stamped, we can trace the hyperlinks in the reverse direction and obtain information cascades. We extracted almost 0.5 million hyperlink cascades from 3.3 million websites from July 2011 till December 2012.
Our aim is to use the hyperlink cascades to infer the time-varying network $G^*(t)$. We then evaluate how many edges \infopath estimates correctly by computing Accuracy,
Precision and Recall for each day.

\begin{figure}[t]
	\centering
	\subfigure[Precision-Recall]{\includegraphics[width=0.22\textwidth]{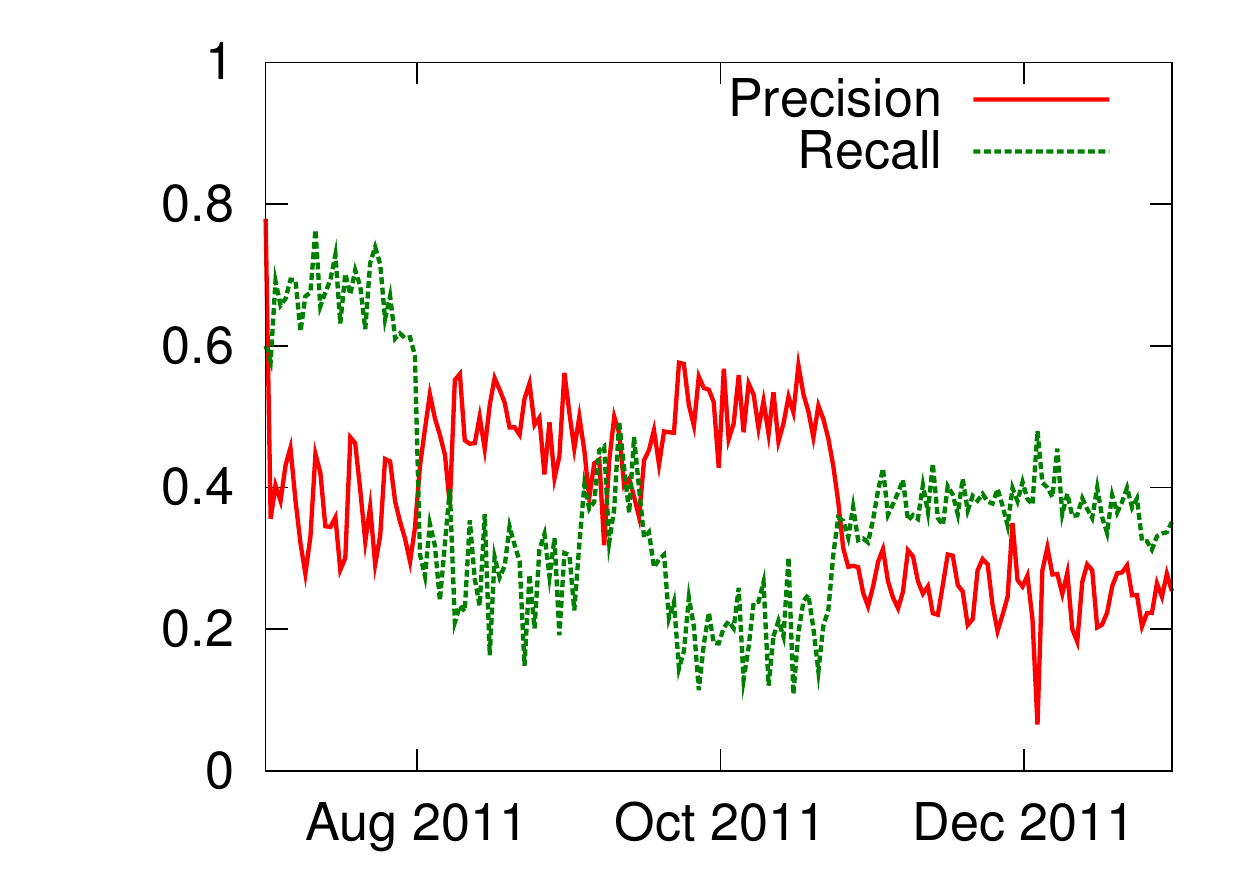} 
	\label{fig:pr-hyperlinks-vs-time}}
	\subfigure[Accuracy]{\includegraphics[width=0.22\textwidth]{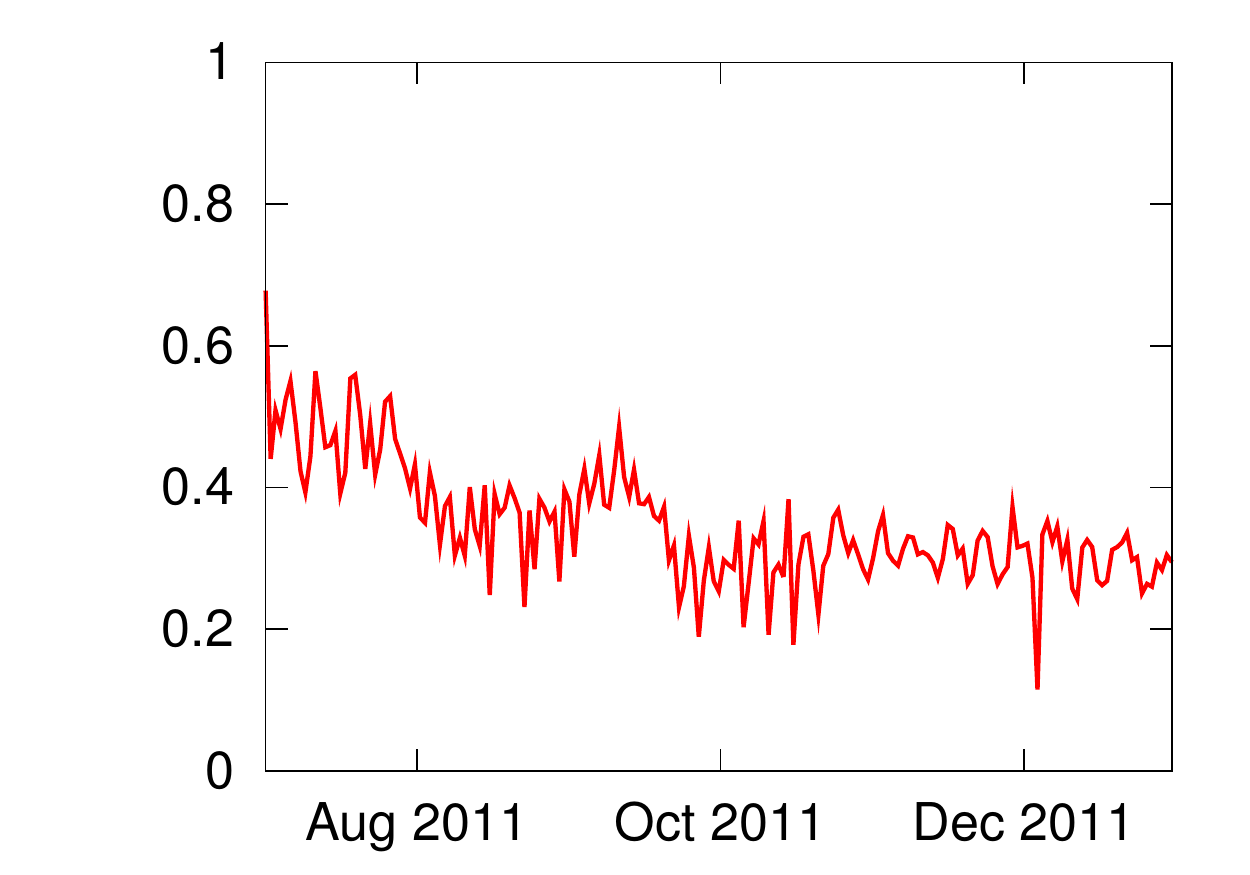} 
	\label{fig:acc-hyperlinks-vs-time}}
 	\vspace{-3mm}
 	\caption{Precision, Recall, and Accuracy of \infopath for a time-varying hyperlink network with 11,461 nodes and 19,915 edges over time, using
	495,655 hyperlink cascades from July 2011 to December 2011.
	} \label{fig:performance-hyperlinks-vs-time}
	\vspace{-3mm}
\end{figure}

Figure~\ref{fig:performance-hyperlinks-vs-time} shows Precision, Recall, and Accuracy over time for a time-varying hyperlink network with 11,461 nodes and 19,915 edges created over time, using 495,655 hyperlink cascades from July 2011 to December 2011. We assume an exponential edge transmission model. We observe weekly periodicity and the overall encouraging performance of around 0.4 to 0.5 for all three performance metrics. 
%Perhaps surprisingly, we observe some variability in the performance of our algorithm over time, with some weekly periodicity. 

\section{Conclusion}
\label{sec:conclusions}
% !TEX root = dynamic-network-inference.tex
All previous network inference algorithms have assumed diffusion networks to be static. Therefore, they have considered the pathways over which information propagates to be 
static over time. In contrast, we developed an algorithm for time-varying network inference, \infopath. Our algorithm provides on-line time-varying estimates of the edges of the network as well as the dynamic edge transmission rates, which allows us to detect how information pathways emerge and vanish over time.

We evaluated our algorithm on synthetic data and demonstrated that \infopath successfully tracks changes in the topology of dynamic networks, provides accurate on-line estimates  of the time-varying edge transmission rates and is also robust across network topologies, edge transmission models and patterns of evolution of edge transmission rates. 

We also run \infopath on real data and investigated how real networks and information pathways evolve over time. 
We found that information pathways over which general recurrent topics propagate remain relatively stable across time. In contrast, major real-world events lead to dramatic 
changes and shifts in the information pathways.
We observed that clusters of mainstream news and blogs often emergence and vanish in matter of days.
We dis\-co\-vered that there is an early greater increase in information transfer among blogs than among mainstream media for news involving general population and social unrest, such as the Libyan civil war, Egyptian revolution, Syria's uprise and the Occupy Wall Street movement.
%Finally, although we found that the amount of mainstream media and blogs among the most central nodes for most topics or news events are comparable, the number of central blogs on some topics or news events grows when there exists an increasing social unrest (\eg, 

Our work also opens various venues for future work. For example, rigorous theoretical analysis of the convergence of our stochastic gradient descent method would provide further insights for its performance. Moreover, we notice that many times the changes in the inferred network structure could be attributed to sudden external real-world events. This opens two interesting questions. How can diffusion network inference be combined with methods for detecting external influence in networks~\cite{seth2012kdd}? And also, how can dynamic network inference be extended for detecting unexpected real-world events based on a stream of documents? Last, many times not only information but also sentiment attached to a piece of information spreads through the network~\cite{miller11flow}. It would be interesting to think about inference of signed networks, where a positive/negative valence of an edge models sentiment relationship between a pair of nodes. Overall, such methods would allow us to improve our understanding of the current landscape of news coverage, the role that news media plays in framing the discussion of important topics, and the evolving ecosystem that news media occupies\-.

\xhdr{Acknowledgements} This research has been supported in part by NSF
IIS-1016909,		    										% NSF with Jon (Sep 2013)
CNS-1010921,		    										% NSF with Madhav (Sept 2015)
IIS-1159679,		    										% NSF Sentiment (Sep 2015)
IIS-1149837,											% NSF CAREER (Dec 2015)
%AFRL FA8650-10-C-7058,									% SRI IARPA (Sep 2014)
DAR\-PA SMISC,													% (Feb 2015)
% DARPA XDATA,													% (?)
% DARPA GRAPHS,													% (Feb 2014)
ARO MURI, %W911NF-12-1-0509	  					% (Sep 2015)
Brown Institute for Media Innovation,	  % Stanford MAGIC (Sept 2013)
Albert Yu \& Mary Bechmann Foundation,	% till Jan 2013
Allyes, 																% till Mar 2013
Boeing, 																% till Dec 2013
Docomo,
Intel,                                  % till Dec 2014
Samsung,																% till Dec 2012
Alfred P. Sloan Fellowship, 					% till June 2013
the Microsoft Faculty Fellowship, 			% till Sept 2013
and Barrie de la Maza Graduate Fellowship.

\nocite{website13}

\bibliographystyle{abbrv}
\bibliography{refs}

\begin{thebibliography}{10}

\bibitem{website13}
\textsc{InfoPath} supporting website: \url{http://snap.stanford.edu/infopath}.

\bibitem{snap}
\textsc{SNAP:} {S}tanford {N}etwork {A}nalysis {P}latform.
  \url{http://snap.stanford.edu}.

\bibitem{aalen2008survival}
O.~Aalen, {\O}.~Borgan, and H.~Gjessing.
\newblock {\em Survival and event history analysis: a process point of view}.
\newblock Springer, 2008.

\bibitem{agarwal2011distributed}
A.~Agarwal and J.~Duchi.
\newblock Distributed delayed stochastic optimization.
\newblock 2011.

\bibitem{bach2011non}
F.~Bach, E.~Moulines, et~al.
\newblock Non-asymptotic analysis of stochastic approximation algorithms for
  machine learning.
\newblock In {\em Advances in Neural Information Processing}, 2011.

\bibitem{bailey75mathematical}
N.~T.~J. Bailey.
\newblock {\em The Mathematical Theory of Infectious Diseases and its
  Applications}.
\newblock Hafner Press, 1975.

\bibitem{blatt2008convergent}
D.~Blatt, A.~Hero, and H.~Gauchman.
\newblock A convergent incremental gradient method with a constant step size.
\newblock {\em SIAM Journal on Optimization}, 18(1):29--51, 2008.

\bibitem{duchi2011ergodic}
J.~Duchi, A.~Agarwal, M.~Johansson, and M.~Jordan.
\newblock Ergodic subgradient descent.
\newblock In {\em Allerton '11: Proc. of the 50th Annual Allerton Conference on
  Communication, Control, and Computing}, 2011.

\bibitem{manuel11icml}
M.~Gomez-Rodriguez, D.~Balduzzi, and B.~Sch\"{o}lkopf.
\newblock {Uncovering the Temporal Dynamics of Diffusion Networks}.
\newblock In {\em ICML '11: Proc. of the 28th International Conference on
  Machine Learning}, 2011.

\bibitem{manuel10netinf}
M.~Gomez-Rodriguez, J.~Leskovec, and A.~Krause.
\newblock {Inferring Networks of Diffusion and Influence}.
\newblock In {\em KDD '10: Proc. of the 16th ACM SIGKDD International
  Conference on Knowledge Discovery in Data Mining}, 2010.

\bibitem{influmax12icml}
M.~Gomez-Rodriguez and B.~Sch\"{o}lkopf.
\newblock {Influence Maximization in Continuous Time Diffusion Networks}.
\newblock In {\em ICML '12: Proc. of the 29th International Conference on
  Machine Learning}, 2012.

\bibitem{multitree12icml}
M.~Gomez-Rodriguez and B.~Sch\"{o}lkopf.
\newblock {Submodular Inference of Diffusion Networks from Multiple Trees}.
\newblock In {\em ICML '12: Proc. of the 29th International Conference on
  Machine Learning}, 2012.

\bibitem{kempe03maximizing}
D.~Kempe, J.~M. Kleinberg, and E.~Tardos.
\newblock Maximizing the spread of influence through a social network.
\newblock In {\em KDD '03: Proc. of the 9th ACM SIGKDD International Conference
  on Knowledge Discovery and Data Mining}, 2003.

\bibitem{lawless1982statistical}
J.~Lawless.
\newblock {\em {Statistical models and methods for lifetime data}}.
\newblock Wiley New York, 1982.

\bibitem{jure06viral}
J.~Leskovec, L.~A. Adamic, and B.~A. Huberman.
\newblock The dynamics of viral marketing.
\newblock In {\em EC '06: Proc. of the 7th ACM conference on Electronic
  commerce}, 2006.

\bibitem{leskovec2009kdd}
J.~Leskovec, L.~Backstrom, and J.~Kleinberg.
\newblock Meme-tracking and the dynamics of the news cycle.
\newblock In {\em KDD '09: Proc. of the 15th ACM SIGKDD International
  Conference on Knowledge Discovery and Data Mining}, 2009.

\bibitem{leskovec2010kronecker}
J.~Leskovec, D.~Chakrabarti, J.~Kleinberg, C.~Faloutsos, and Z.~Ghahramani.
\newblock {Kronecker graphs: An approach to modeling networks}.
\newblock {\em The Journal of Machine Learning Research}, 11:985--1042, 2010.

\bibitem{nowell08letter}
D.~Liben-Nowell and J.~Kleinberg.
\newblock Tracing the flow of information on a global scale using {I}nternet
  chain-letter data.
\newblock {\em Proc. of the National Academy of Sciences}, 105(12):4633--4638,
  2008.

\bibitem{miller11flow}
M.~Miller, C.~Sathi, D.~Wiesenthal, J.~Leskovec, and C.~Potts.
\newblock Sentiment flow through hyperlink networks.
\newblock In {\em ICWSM '11: Proc. of the AAAI International Conference on
  Weblogs and Social Media}, 2011.

\bibitem{meyers10netinf}
S.~Myers and J.~Leskovec.
\newblock {On the Convexity of Latent Social Network Inference}.
\newblock In {\em NIPS '10: Advances in Neural Information Processing Systems},
  2010.

\bibitem{myers12clash}
S.~Myers and J.~Leskovec.
\newblock Clash of the contagions: Cooperation and competition in information
  diffusion.
\newblock In {\em ICDM '12: Proc. of the 12th IEEE International Conference on
  Data Mining}, 2012.

\bibitem{seth2012kdd}
S.~Myers, C.~Zhu, and J.~Leskovec.
\newblock Information diffusion and external influence in networks.
\newblock In {\em KDD '12: Proc. of the 18th ACM SIGKDD international
  conference on Knowledge discovery and data mining}, 2012.

\bibitem{nemirovski2009robust}
A.~Nemirovski, A.~Juditsky, G.~Lan, and A.~Shapiro.
\newblock Robust stochastic approximation approach to stochastic programming.
\newblock {\em SIAM Journal on Optimization}, 19(4):1574, 2009.

\bibitem{netrapalli12}
P.~Netrapalli and S.~Sanghavi.
\newblock Finding the graph of epidemic cascades.
\newblock In {\em Proc. of the 12th ACM SIGMETRICS/Performance}, 2012.

\bibitem{prakash2012winner}
B.~Prakash, A.~Beutel, R.~Rosenfeld, and C.~Faloutsos.
\newblock Winner takes all: competing viruses or ideas on fair-play networks.
\newblock In {\em Proc. of the 21st International Conference on World Wide
  Web}, pages 1037--1046, 2012.

\bibitem{robbins1951stochastic}
H.~Robbins and S.~Monro.
\newblock A stochastic approximation method.
\newblock {\em The Annals of Mathematical Statistics}, pages 400--407, 1951.

\bibitem{rogers95diffusion}
E.~M. Rogers.
\newblock {\em Diffusion of Innovations}.
\newblock Free Press, New York, fourth edition, 1995.

\bibitem{romero11twitter}
D.~M. Romero, B.~Meeder, and J.~Kleinberg.
\newblock Differences in the mechanics of information diffusion across topics:
  idioms, political hashtags, and complex contagion on twitter.
\newblock In {\em WWW '11: Proc. of the 20th International Conference on World
  Wide Web}, 2011.

\bibitem{roux2012stochastic}
N.~Roux, M.~Schmidt, and F.~Bach.
\newblock A stochastic gradient method with an exponential convergence rate for
  strongly-convex optimization with finite training sets.
\newblock {\em Advances in Neural Information Processing}, 2012.

\bibitem{snowsill2011refining}
T.~Snowsill, N.~Fyson, T.~De~Bie, and N.~Cristianini.
\newblock Refining causality: who copied from whom?
\newblock In {\em KDD '11: Proc. of the 17th ACM SIGKDD international
  conference on Knowledge discovery and data mining}, 2011.

\bibitem{wallinga04epidemic}
J.~Wallinga and P.~Teunis.
\newblock Different epidemic curves for severe acute respiratory syndrome
  reveal similar impacts of control measures.
\newblock {\em American Journal of Epidemiology}, 160(6):509--516, 2004.

\bibitem{yang2010}
J.~Yang and J.~Leskovec.
\newblock Modeling information diffusion in implicit networks.
\newblock In {\em ICDM '10: Proc. of the IEEE International Conference on Data
  Mining}, 2010.

\bibitem{yang2010patterns}
J.~Yang and J.~Leskovec.
\newblock {Patterns of Temporal Variation in Online Media}.
\newblock In {\em WSDM '11: ACM International Conference on Web Search and Data
  Mining}, 2011.

\end{thebibliography}

\end{document}